\documentclass[%
 reprint, superscriptaddress,
 amsmath,amssymb,
 aps,
]{revtex4-2}

\usepackage{graphicx}% Include figure files
\usepackage{dcolumn}% Align table columns on decimal point
\usepackage{bm}% bold math
\usepackage{siunitx}
\usepackage[normalem]{ulem}
\usepackage[hidelinks]{hyperref}
\usepackage{orcidlink}

\usepackage{xcolor}
\definecolor{amber}{rgb}{1,0.49,0}

\newcommand{\editor}[2]{%
	\expandafter\newcommand\csname #1note\endcsname[1]{%
		\textcolor{#2}{(\textbf{#1:} ##1)}}%
	\expandafter\newcommand\csname #1\endcsname[1]{%
		\textcolor{#2}{##1}}%
	\expandafter\newcommand\csname #1cancel\endcsname[1]{%
		\textcolor{#2}{\sout{##1}}}%
	\expandafter\newcommand\csname #1change\endcsname[2]{%
		\textcolor{#2}{\sout{##1} ##2}}%
	\expandafter\newcommand\csname #1prov\endcsname[2]{%
		\textcolor{#2}{[(##1) (##2)]}}%
	\newenvironment{#1text}{\color{#2}}{\color{black}}
}

\editor{SB}{amber}
\editor{FP}{blue}
\begin{document}

\preprint{APS/123-QED}

\title{Glassy Dynamics from First-Principles Simulations }% Force line breaks with \\

\author{Florian Pabst\,\orcidlink{0000-0001-9331-5172}}
\email{fpabst@sissa.it}
\affiliation{%
 SISSA – Scuola Internazionale Superiore di Studi Avanzati, Trieste (Italy)
}%

\author{Stefano Baroni\,\orcidlink{0000-0002-3508-6663}}%
 \affiliation{%
 SISSA – Scuola Internazionale Superiore di Studi Avanzati, Trieste (Italy)
}%
\affiliation{CNR-IOM, Istituto dell'Officina dei Materiali, SISSA unit, Trieste (Italy)}

\date{\today}% It is always \today, today,

\begin{abstract}
The microscopic understanding of the dramatic increase in viscosity of liquids when cooled towards the glass transition is a major unresolved issue in condensed matter physics. 
Here, we use machine learning methods to accelerate molecular dynamics simulations with first-principles accuracy for the glass-former toluene. We show that the increase in viscosity is intimately linked to the increasing number of dynamically correlated molecules $N^*$. While certain hallmark features of glassy dynamics, like physical aging, are linked to $N^*$ as well, others, like relaxation stretching, are not. 

\end{abstract}

\maketitle
%\tableofcontents

Glasses are ubiquitous in our everyday lives. For instance, the water bottle on your desk might be made from silicate glass, commonly referred to as \emph{glass} in everyday language, or from plastic, which is a type of polymeric glass. Unlike crystalline materials, glasses lack long-range molecular order, resembling liquids, and there is no phase transition in the conventional sense between the liquid and glass state. Instead, as a liquid cools, its viscosity continuously increases until it becomes so high that we consider it a solid---and call it glass. But what is the microscopic origin of the order-of-magnitude difference in viscosity between high and low temperatures? Despite decades of research, there is no comprehensive, accepted theory to answer this question \cite{berthier2011theoretical}. Phenomenologically, the temperature evolution of the viscosity $\eta$ can be described by 
\begin{equation}
    \eta = \eta_0 \exp \left( \frac{E^*(T)}{RT} \right)
  \label{eq_arrh}
\end{equation}
where $\eta_0$ is some high temperature limiting value of the viscosity, $E^*(T)$ a temperature-dependent activation energy, and R is the gas constant. It is commonly assumed that in the liquid state at high temperatures $E^*$ is independent of temperature, i.e., Eq.~\eqref{eq_arrh} is the Arrhenius equation in this case. This equation can be rationalized by assuming that the molecules have to overcome an energy barrier $E^*$ in order to move, thus enabling viscous flow.
However, at lower temperatures, it is found for most liquids that the viscosity increases more strongly than the Arrhenius equation would suggest, which can be formally accounted for by an increase in the activation energy $E^*(T)$ upon cooling. A common picture for this increase is that the molecules start to move in a concerted way at lower temperatures, leading to an effectively higher energy barrier which has to be overcome collectively by a growing number of molecules. Although this idea has been around in various forms since the seminal work of Adam-Gibbs \cite{adam1965temperature}, its validity as the driving mechanism for the observed increase of $E^*(T)$ at low temperature is still vividly debated today \cite{pica2024local}.

The difficulty here is to define and determine the number of dynamically correlated molecules. In order to address this difficulty, it is convenient to define (the self-part of a) relaxation function as \cite{Parisi1997,harrowell2011length}:
\begin{equation}
    Q(t) = \frac{1}{N} \sum\limits_{i=1}^{N}w\left(|\bm r_i(t_0) - \bm r_i(t+t_0)| \right),
    \label{eq:RelaxationFunction}
\end{equation}
where $\bm r_i(t)$ is the position of the center of mass of the $i$-th molecule at time $t$ and $w(r)$ is a window function such that $w(r) = 1$ if $r \leq a$ and $0$ otherwise, $a$ being a distance larger than the vibrational amplitude of the molecules (usually $a \approx 1/3$ of the intermolecular distance is used). At small time, $t\to 0$, $Q(t)$ tends to 1, whereas at large times, $t\to\infty$, it tends to 0. 
At intermediate times, the shape of $Q(t)$ depends more on $t_0$ the 
larger the number of dynamically correlated molecules $N^*$ (see SI). It seems therefore natural to define $N^*$ as to be related to the maximum of the variance of $Q(t)$ like: 
\begin{equation}
    N^*(T)\propto \frac{V}{T} \max \limits_{t} \Delta Q^2(t), \label{eq:N}
\end{equation} 
where $\Delta Q(t)^2= \bigl \langle Q^2(t) \bigr \rangle - \bigl \langle Q^{}(t) \bigr \rangle ^2$ and $V$ is the volume of the sample \cite{harrowell2011length}.

$\Delta Q^2$ is not directly measurable by experiments, but can be accessed approximately by nonlinear dielectric measurements \cite{crauste2010evidence}. However, although these measurements---and others in similar directions \cite{albert2016fifth,berthier2005direct}---all obtain a value for $N^*$ or a correlation length $L^* \propto \sqrt[3]{N^*}$ which grows upon cooling, we are aware of only one study that found a direct connection between $N^*$ and $E^*(T)$ for four structural glass-former \cite{bauer2013cooperativity}.

In computer simulations, $\Delta Q^2$ is easily accessible, and increasing values of $N^*$ upon cooling have also been found \cite{karmakar2009growing}. However, like most of the computer simulations on glassy dynamics, these studies focus on binary Lennard-Jones model liquids. Although invaluable insight into the dynamics of glassy liquids has been gained in this way \cite{sastry1998signatures}, their relevance to the understanding of real liquids can only be qualitative. Moreover, it has recently been shown that the polydispersity of particle sizes necessary to avoid crystallization in these models might drastically change the dynamics as compared to a mono-component liquid \cite{pihlajamaa2023influence}. 

\begin{figure}[htb!]
\includegraphics[width = 0.48\textwidth]{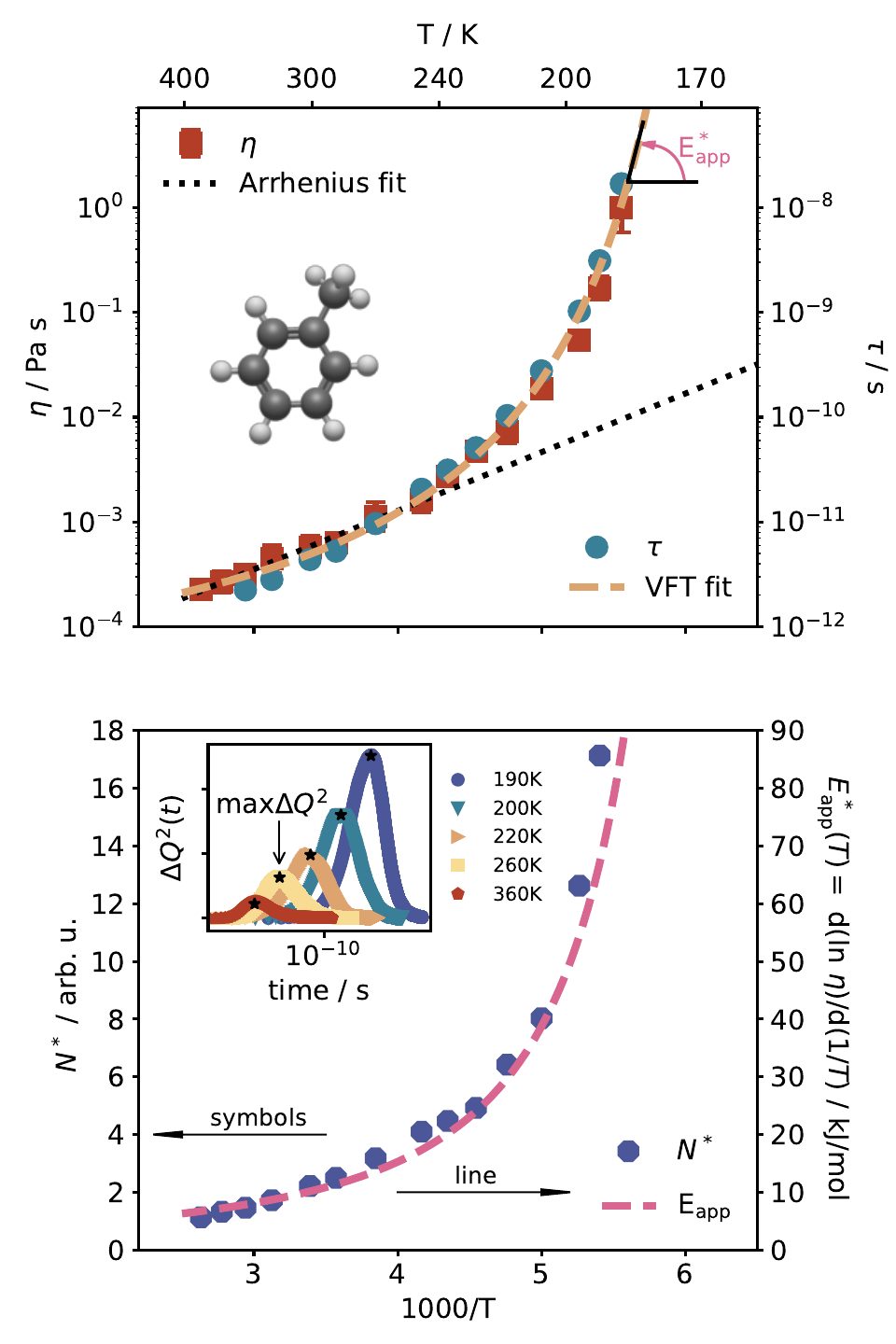}
\caption{\label{fig:arrh} Top: Viscosity and rotational correlation times as a function of inverse temperature. Bottom: Number of dynamically correlated molecules $N^*(T)$ (filled octagons) as obtained from $\Delta Q(t)$ shown in the inset. The apparent activation energy $E^*_{\rm app}$ obtained from the slope of the VFT curve in the upper panel is included as a dashed line, showing the proportionality $N^*(T) \propto E^*_{\rm app}(T)$.  }
\end{figure}

The purpose of this paper, therefore, is to study glassy dynamics, and especially the reason for the increase of $E^*(T)$ upon cooling, on as realistic a liquid model as possible. We choose toluene (see Fig.~\ref{fig:arrh} for a sketch of the chemical structure), as it is a small molecule with only two atom types and experimentally well studied. Highly accurate simulations can nowadays be performed using ab-initio molecular dynamics (AIMD) methods based on density-functional theory (DFT), but due to the high computational costs of these methods, the accessible time scales (usually a few hundred picoseconds at most) are too short to reach the above-mentioned regime where $E^*$ is expected to become temperature dependent. Therefore, we use a machine-learning approach to train a neural network potential on DFT data \cite{zhang2020dp,giannozzi2009quantum}, allowing us to run simulations with first-principles accuracy, but roughly a factor 1000 longer than AIMD would allow. This makes it possible to reach the necessary temperature and time ranges, with the longest trajectory of this work being \SI{0.5}{\micro s}. All data shown here are from simulations in the microcanonical ensemble on a 512 molecules (7680 atoms) system, except for the lowest temperature (\SI{180}{K}) where only 32 molecules are used due to the high computational cost, leading to poorer statistics due to the reduced number of molecules. We did not find any size dependence of the shown quantities between 32 and 8192 molecules, which is demonstrated in the supporting information (SI), where also all details of the simulations can be found. 

In the top panel of Fig.~\ref{fig:arrh} we report the dependence of the viscosity, $\eta$, on inverse temperature. The structural relaxation time $\tau$ is connected to $\eta$ via the high-frequency shear modulus $G_{\infty}$, where the temperature dependence of the latter is negligible: $\tau(T) = \eta(T)/G_{\infty}$ An independent measure of $\tau$ can be obtained from the rotational correlation function of the molecules, as detailed below. These relaxation times $\tau$ are also displayed in the top panel of Fig.~\ref{fig:arrh}.  At low temperatures, the deviation of the data from the Arrhenius law fitted to the high-temperature portion (dotted line) is obvious, indicating that we indeed reach the glassy regime where $E^*(T)$ is no longer constant. The data could be fitted over the whole temperature range with the commonly used empirical Vogel-Fulcher-Tamann (VFT) equation,  $\eta(T) = \eta_{\rm VFT} \frac{D}{T-T_0}$, where $\eta_{\rm VFT}$, $D$ and $T_0$ are fitting parameters. From this fit, we calculate an ``apparent'' activation energy  $E^*_{\rm app}(T) \approx R\, \text{d}(\ln\eta)/\text{d}(1/T)$, i.e., the activation energy an Arrhenius law would have at that temperature, which can thus be visualized as the slope of the VFT curve at the respective temperature.  

In the bottom panel of Fig.~\ref{fig:arrh}, $N^*(T)$ is shown as a function of inverse temperature as obtained via Eq.~\eqref{eq:N} from the $\Delta Q^2(t)$ data shown in the inset.  
Included in that panel (scale on the right y-axis) is $E^*_{\rm app}(T)$ as a dashed line. Since $N^*$ can only be determined modulo an unknown constant multiplicative factor, this factor is chosen in such a way that overlap between $N^*(T)$ on the left y-axis and $E^*_{\rm app}(T)$ on the right y-axis is achieved.

The striking finding is that $N^*(T) \propto E^*_{\rm app}(T)$ to very good approximation as seen by the accordance of the points and the dashed line in the lower panel of Fig.~\ref{fig:arrh}. As mentioned above, such a proportionality was reported in an experimental study for four molecular glass-formers before \cite{bauer2013cooperativity}. However, since the experiment did not measure $\Delta Q^2(t)$ directly but a quantity which is sometimes discussed to be unrelated to $N^*(T)$ \cite{diezemann2012nonlinear}, our finding based on the direct determination of $\Delta Q^2(t)$ strongly supports the experiments of Ref.  \citenum{bauer2013cooperativity}. Moreover, our simulations cover a much larger range in both temperature and $N^*(T)$, which makes the proportionality $N^*(T) \propto E^*_{\rm app}(T)$ even more evident. Additionally, since our simulations extend up to the boiling point of toluene $T_{\rm bp}= 384~\mathrm{K}$, we can make the reasonable assumption that $N^*(380K) = 1$, i.e., molecules are correlated only with itself near the boiling point, which is backed up by the discussion of Fig.~\ref{fig:mobile} below. The scaling of the left y-axis in Fig.~\ref{eq_arrh} is based on this assumption, giving a maximum of $\approx 17$ correlated molecules in our simulations. This number is in accordance with the fact that we do not see finite-size effects in our simulations down to 32 molecules. Applying the same method to the experimental value $E^*_{\rm app}(T_g)$ \cite{do1997dielectric} also allows us to roughly approximate the number of dynamical correlated molecules at the glass transition temperature $T_g=\SI{117}{K}$, and we find $N_{\rm exp.}^*(T_g) \approx 43$.  Finite-size effects in experiments on a different glass-former with basically identical molecular weight as toluene indicate that the magnitude of $N^*$ approximated here is very reasonable \cite{uhl2019glycerol}. We note that the $N^*$ values do not support the existence of an Arrhenius-regime, i.e. $E^*_{\rm app}(T) =$ const., which would result in a horizontal slope in the lower part of Fig.~\ref{fig:arrh}.
\begin{figure}[htb!]
\includegraphics[width = 0.48\textwidth]{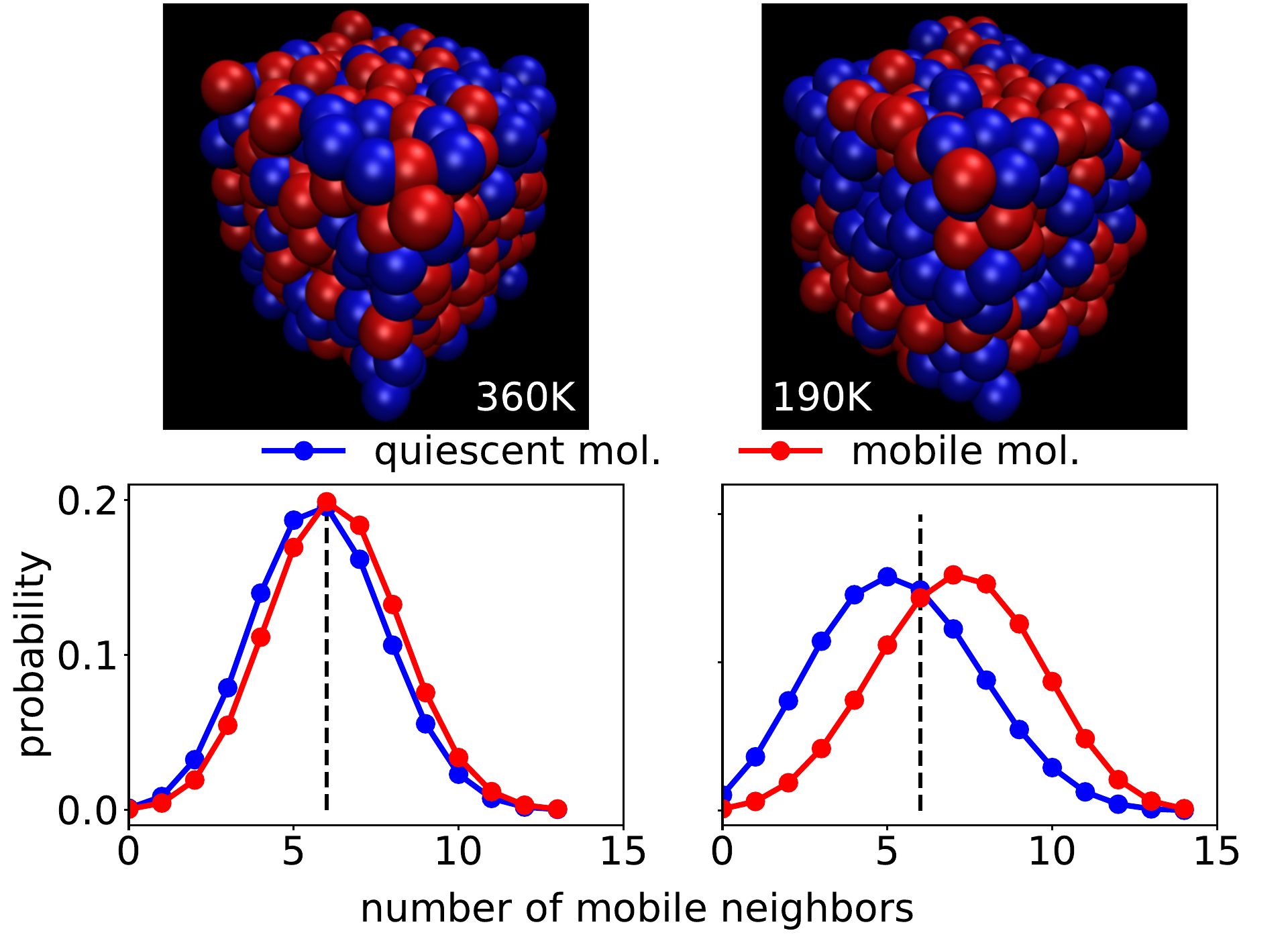}
\caption{\label{fig:mobile} Check for dynamical heterogeneity. At \SI{360}{K} the snapshot shows rather a random arrangement of mobile (red) and quiescent (blue) molecules. At \SI{190}{K} clusters of red and clusters of blue spheres (i.e. centers of mass of the toluene molecules) can be seen. At \SI{190}{K} the number of mobile neighbors is higher for a mobile molecule than for a quiescent molecule, while it is similar at \SI{360}{K}.}
\end{figure}

It is also possible to visualize the space-time dynamical implications of different values of $N^*$. This is done for two cases representative for the high- and low-temperature regimes in Fig.~\ref{fig:mobile}, where the same criterion as in Eq.~\eqref{eq:RelaxationFunction} is used to discriminate between molecules of different mobility: 
we label as \emph{mobile} a molecule that has traveled a distance larger than $a=\SI{2}{\angstrom}$ at a time when half of the molecules of the sample have done so, and as \emph{quiescent} the others. In the upper part of Fig.~\ref{fig:mobile}, the centers of mass of mobile and quiescent molecules are depicted as red and blue spheres, respectively.

It is clear that at \SI{360}{K} mobile and quiescent molecules are distributed quite randomly, while at \SI{190}{K} clusters of red and clusters of blue spheres can be clearly identified. This is a manifestation of a hallmark feature of glassy dynamics called \emph{spatial dynamic heterogeneity} \cite{ediger2000spatially}, and a direct consequence of the increasing number of dynamically correlated molecules at low temperatures. In the lower part of Fig.~\ref{fig:mobile} this behavior is quantified by counting how many mobile neighbors a mobile (red circles) or a quiescent (blue circles) molecule has. While this number is basically identical at \SI{360}{K}, it is more probable to find a mobile molecule in the neighborhood of a mobile molecule than in the neighborhood of a quiescent one at \SI{190}{K}. The same information is contained in the partial radial distribution functions $g(r)$ shown in the SI. We note that more subtle spatial-dynamic correlations can be resolved with a recently proposed method \cite{zhang2020revealing}: At a distance from the reference molecule corresponding to dips in $g(r)$, it is more likely to find mobile molecules than to find quiescent ones, while the opposite is true for peaks in $g(r)$ (see SI for details). This means that we do not see rigid cooperatively rearranging regions like proposed e.g. in the Adam-Gibbs theory \cite{adam1965temperature}, but a delicate interplay between dynamic and structural correlations. 

\begin{figure}[htb!]
\includegraphics[width = 0.48\textwidth]{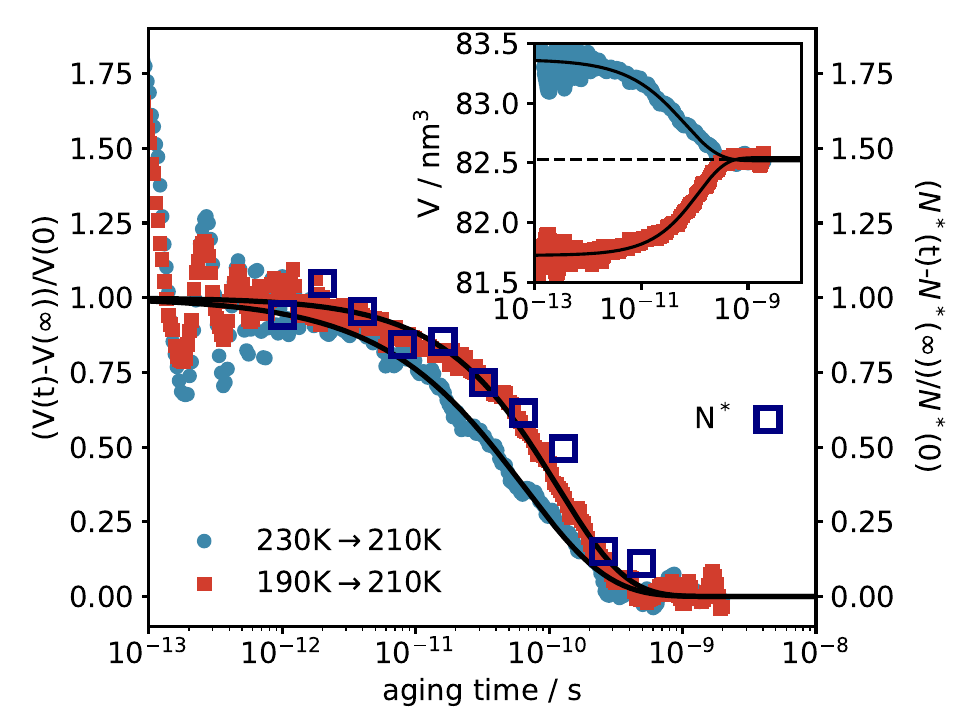}
\caption{\label{fig:aging} The box volume after a temperature jump of $ \Delta T = \pm 20K$ and $N^*$ after the temperature down jump exhibit very similar aging time dependence.}
\end{figure}

Another hallmark feature of glassy dynamics is physical aging, i.e., the slow evolution of the properties of a liquid---like the volume---towards equilibrium after a jump in temperature of $\Delta T$. The approach to equilibrium is dependent on the thermal history, meaning that it does not only depend on the final temperature but also on the 
temperature 
before the jump. This is shown in Fig.~\ref{fig:aging}, where in the inset (the moving average of) the box volume after a temperature jump of $ \Delta T = \pm 20K$ is shown. In the main panel the same data is normalized to better visualize the difference between an upward and a downward jump. 

It can be seen that the volume evolution after the downward jump is more stretched than after the upward jump. This behavior is well-known from experiments \cite{riechers2022predicting}. Empirical models exist to rationalize this finding based on the idea that the instantaneous relaxation time during aging not only depends on the actual temperature but also on the so-called fictive temperature, which is the temperature the system with the present structure would have when in equilibrium \cite{riechers2022predicting}. Although an underlying assumption of this model has recently been confirmed by experiment \cite{bohmer2024time},
a thorough theoretical understanding of the aging dynamics is still lacking, which is directly connected to the absence of a theory for the dynamics of glassy materials at equilibrium. A growing $N^*$ after a temperature downward jump has been observed by experiment \cite{brun2012evidence} and simulation \cite{parsaeian2008growth}, but to the best of our knowledge, a direct connection to the aging time dependence of the sample volume has not been reported so far. Included in Fig.~\ref{fig:aging} is the evolution of the normalized number of correlated molecules after the \SI{20}{K} down-jump, obtained from the maximum of $\Delta Q^2(t)$, dependent on aging time. Although the uncertainty of these values is much higher than for the equilibrium simulations above due to poorer statistics (see SI for details), it can be clearly seen that the evolution of $N^*$ approximately resembles the one of the volume. This shows that also physical aging is closely connected to the number of dynamically correlated molecules. 

It could almost considered common sense that the shape of correlation functions, e.g. of density fluctuations or rotational dynamics, in glassy materials is non-exponential due to the presence of dynamical heterogeneity. The idea is that at high temperatures in the normal liquid regime, the relaxation decay is exponential ($C = \exp(-t/\tau)$), while it becomes broadened ($C = \exp(-(t/\tau)^{\beta})$, with $0< \beta <1 $) at lower temperatures due to a broadened distribution of relaxation times $\tau$ caused by dynamical heterogeneity. However, this picture is problematic for several reasons: Relaxation stretching ($\beta<1$) is experimentally still observed at temperatures close to the boiling point \cite{schmidtke2014relaxation}. Time-temperature-superposition (TTS), i.e. the invariance of the $\beta$ parameter with temperature has been reported \cite{olsen2001time}, which is hard to reconcile with dynamical heterogeneity supposed to increase with lowering temperature. Finally, a prevalence of $\beta \approx 0.5$ has been reported for various kinds of liquids \cite{pabst2021generic}, which would imply a similar dynamic heterogeneity for chemically largely different substances. 

\begin{figure}
\includegraphics[width = 0.48\textwidth]{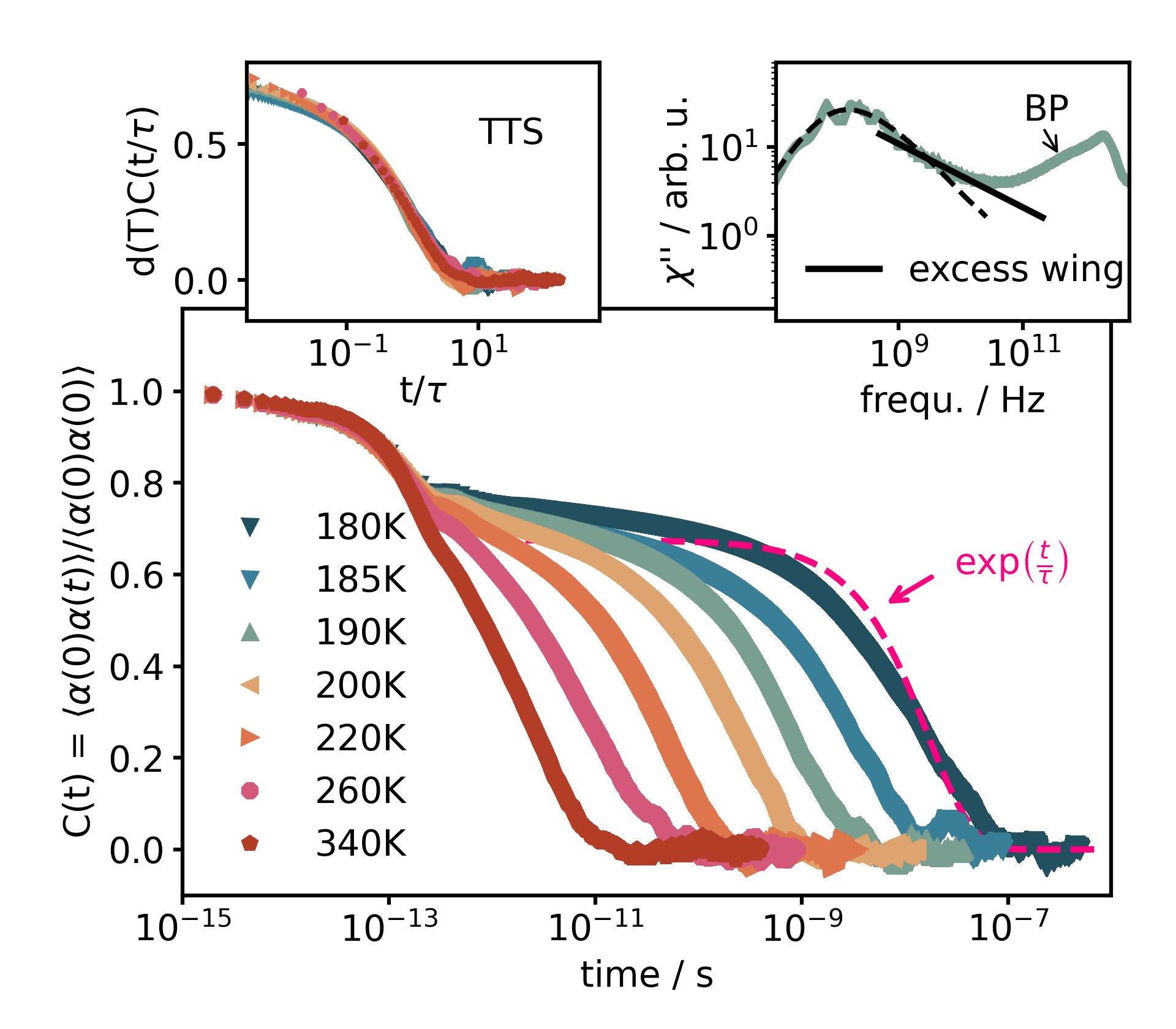}
\caption{\label{fig:LS} Rotational correlation functions for selected temperatures exhibiting non-exponential shape. Left inset: Demonstration of time-temperature-superposition (TTS). Right inset: Frequency domain representation of the data showing clearly an excess wing and the boson peak.}
\end{figure}

In Fig.~\ref{fig:LS} we show rotational correlation functions calculated from the off-diagonal elements of the polarizability tensor $\alpha(t)$, as could also be measured by depolarized light scattering experiments \cite{pabst2022intensity} (see SI for details). 
The dashed magenta line shows an exponential decay and it is clear that the data is more stretched, i.e. $\beta<1$. However, this stretching is independent of temperature up to close to the boiling point, as shown in the left inset of Fig.~\ref{fig:LS} where the data is scaled by $\tau(T)$ on the x-axis and for some temperatures by a factor $d(T)$ close to unity on the y-axis to achieve good overlap of all curves, highlighting the common shape. Since $N^*$, as well as the dynamical heterogeneity, increases strongly in this temperature range, it can be concluded that the relaxation stretching is not related to either of them. TTS is also found for translational motions, as shown in the SI. We note in passing that we also do not find any sign of violation of the Stokes-Einstein relation, i.e., the proportionality of the inverse diffusion coefficient and the viscosity, as often believed to arise as a consequence of dynamical heterogeneity \cite{kawasaki2017identifying}. 

In the right inset of Fig.~\ref{fig:LS} we show the data at one temperature in the frequency domain in the susceptibility representation 
$\chi''(\omega) = \omega \int\limits_{-\infty}^{\infty} C(t) e^{i\omega t} \text{d}t $
to better visualize two more prominent features of glassy dynamics: On the one hand, the excess wing, highlighted by the solid black line, which was very recently observed for the first time in a molecular dynamics simulation of a model liquid \cite{guiselin2022microscopic}. It is sometimes considered to be the high-frequency flank of a secondary relaxation \cite{schneider2000excess}, ubiquitous in glassy materials, but still poorly understood. On the other hand, the boson peak (BP) is found, marked by the arrow, usually ascribed to an excess density of vibrational states and responsible for the low-temperature anomaly of the heat capacity of glasses \cite{shintani2008universal}.

In conclusion, we have shown in the specific example of toluene that it is now possible to perform molecular dynamics simulation with first-principles accuracy in the temperature range where all features of glassy dynamics can be clearly observed. This was possible by using machine learning methods to train a neural network potential on DFT data. It opens up the possibility to study all aspects of glassy dynamics and to test theories concerned with the glass transition thoroughly in the future on realistic liquids. Furthermore, rotational spectra, which play a crucial role in experimental studies of glass forming liquids \cite{lunkenheimer2000glassy}, but are not accessible in Lennard-Jones liquids, can be calculated from our simulations with high accuracy due to the available electronic structure. Against common wisdom, we found that the shape of the structural relaxation in these spectra does not broaden with increasing dynamical heterogeneity. Also, there is no transition between a high-temperature liquid and a low-temperature glassy regime, found in simulations employing simple model liquids \cite{sastry1998signatures}. Instead, quantities like the apparent activation energy $E^*_{\rm app}$ evolve smoothly with temperature starting from the boiling point. Remarkably, $E^*_{\rm app}$ is proportional to the number of dynamically correlated molecules $N^*$, highlighting the intimate connection between the order-of-magnitude increase of the viscosity and the moderate increase of $N^*$ upon cooling. This finding is in accordance with experimental results, but was not reported for simulations before and challenges theories explaining the increasing activation energy without reference to an increasing number of dynamically correlated molecules \cite{dyre2006colloquium}.

\section*{Acknowledgments}
The authors are grateful to Alfredo Fiorentino, Cesare Malosso, Enrico Drigo and Paolo Pegolo for fruitful discussions.
This work was partially supported by the European Commission through the \textsc{MaX} Centre of Excellence for supercomputing applications (grant number 101093374), by the Italian MUR, through the PRIN project ARES (grant number 2022W2BPCK), and by the Italian National Centre for HPC, Big Data, and Quantum Computing (grant number CN00000013), funded through the \emph{Next generation EU} initiative. 

\bibliography{bib}% Produces the bibliography via BibTeX.

\end{document}

\title{Glassy Dynamics from First-Principles Simulations \\ SUPPORTING INFORMATION}% Force line breaks with \\

\author{Florian Pabst}
\email{fpabst@sissa.it}
\affiliation{%
 SISSA – Scuola Internazionale Superiore di Studi Avanzati, Trieste (Italy)
}%

%\email{fpabst@sissa.it}
\author{Stefano Baroni}%
 \affiliation{%
 SISSA – Scuola Internazionale Superiore di Studi Avanzati, Trieste (Italy)
}%
\affiliation{CNR-IOM, Istituto dell'Officina dei Materiali, SISSA unit, Trieste (Italy)}


\maketitle

\section{Neural Network Potential} \label{sec:NNP}
\begin{figure}[h!]
    \centering
    \includegraphics[width=0.5\textwidth]{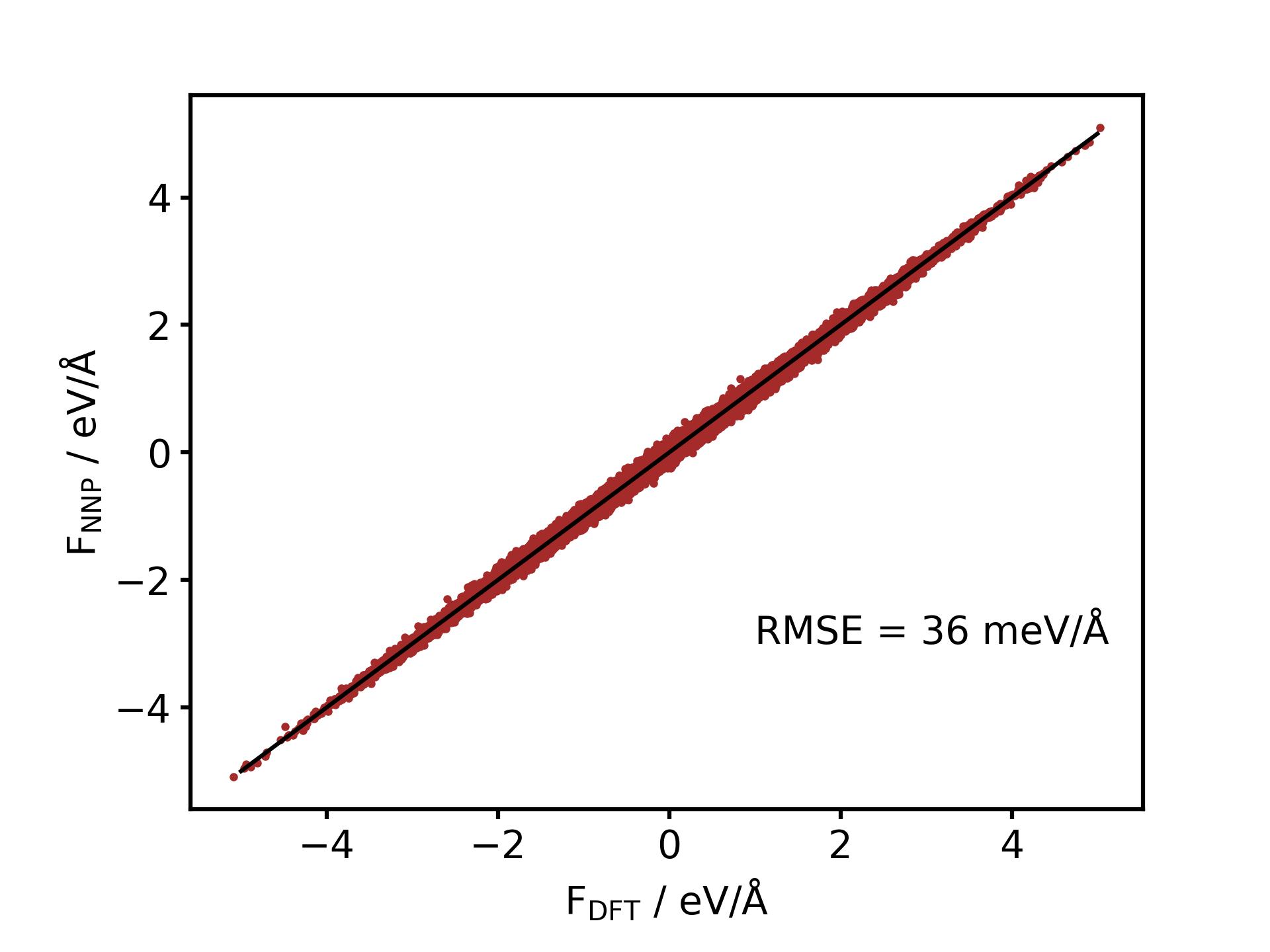}\
    \caption{Comparison of forces predicted by the NNP and from DFT calculations.}
    \label{fig:forces}
\end{figure}

The Neural-Network Potential (NNP) used in this work is trained via a recently
proposed “on-the-ﬂy” learning procedure called Deep Potential Generator (DP-GEN) \cite{zhang2020dp}. The workflow consists of three steps which are repeated until convergence of the learning procedure is achieved. The first step is the training of four NNPs, differing only by the initialization seed, which is done in the very first iteration on data taken from short ab-initio simulations. These NNPs are then used in the second step to run short simulations in the NVT or NPT ensemble with temperatures up to \SI{480}{K} and pressures up to \SI{10}{kbar}. The difference in the forces predicted by the four NNPs is used as a criterion for selecting a snapshot for inclusion in the next training iteration. Forces and energies are calculated for the selected snapshots in the third step using density functional theory (DFT) with \qe\ \cite{giannozzi2009quantum}.
The RPBE functional and D3(BJ) dispersion correction is used for these calculations \cite{hammer1999improved,grimme2011effect}, as well as optimized norm-conserving Vanderbilt pseudo-potentials \cite{hamann2013optimized} and a kinetic energy cutoff of \SI{85}{Ry}.
The training of the final NNP is done on 9558 configurations of 16 molecules for 7 million steps. The cutoff radius was set to \SI{7}{\angstrom} and the size of the embedding and ﬁtting nets is (25, 50, 100) and (240, 240, 240), respectively. After the training, the NNP was compressed to increase the speed of the simulations \cite{lu2022dp}.
Fig.~\ref{fig:forces} shows predicted forces of the NNP for 1040 configurations obtained during the actual production runs with 32 molecules, and therefore not contained in the training set, versus DFT forces. The average error on the forces is \SI{36}{meV\angstrom^{-1}}, which is well below the threshold of \SI{50}{meV\angstrom^{-1}} commonly considered as accurate for a DeepMD potential.
\newline

\clearpage

\section{Simulation Details}
\begin{figure}[h!]
    \centering
    \includegraphics[width=0.5\textwidth]{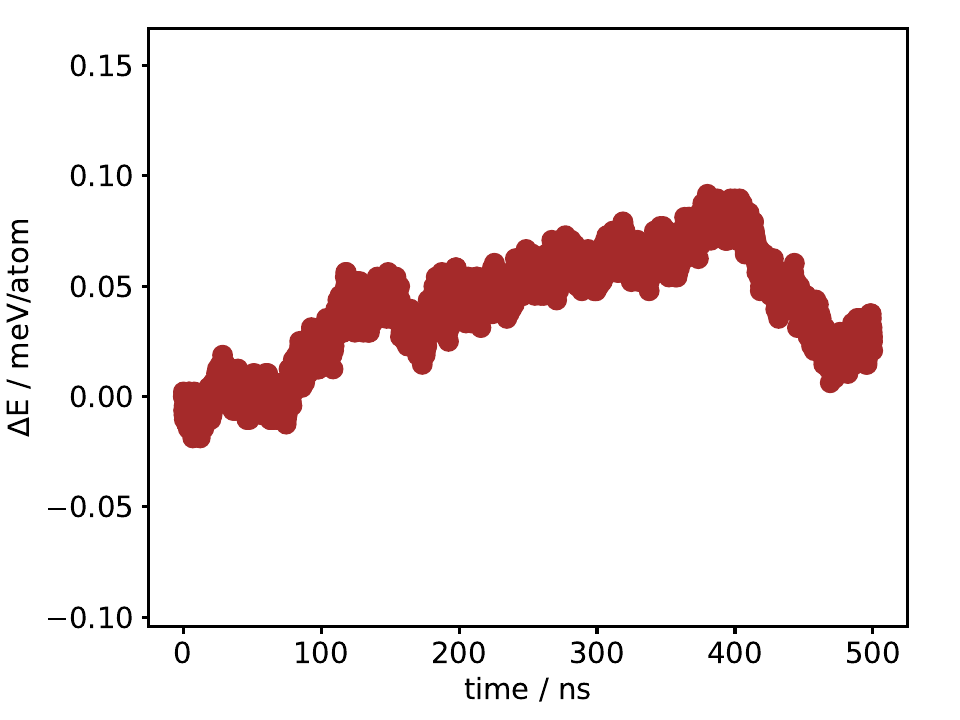}\
    \caption{Drift of the total energy during the longest NVE simulation (32 molecules) of this work. Excellent energy stability is observed. }
    \label{fig:energy}
\end{figure}

All simulations were performed using the LAMMPS code interfaced with the DeepMD-kit \cite{thompson2022lammps,zeng2023deepmd}. Equilibration runs with a time step of \SI{1}{fs} were performed in the NVT ensemble, where the density was set to the (linearly extrapolated) experimental density, which is basically identical to the density obtained in NPT simulations (slight deviations less than 3\% are seen at the highest temperatures, see Fig.~\ref{fig:sfac-density} below). After equilibration, production runs are performed on 32, 512 or 8192 toluene molecules in the NVE ensemble using a time step of \SI{0.5}{fs}. For each temperature of the 32 and 512 molecules simulations 8 simulations are started from independent configurations of the equilibration run (separated at least by $3\tau$, where $\tau$ is the relaxation time at the respective temperature) and all reported quantities are the average over these 8 runs. For the 8192 molecules (122880 atoms) simulations, only one run was performed at each temperature due to the high computational cost.   
Equilibration and production were both run for approximately 100 times longer than the relaxation time $\tau$ at the respective temperatures. This criterion was somewhat relaxed at low temperatures, with no simulation being shorter than $30\tau$.
In Fig.~\ref{fig:energy} the excellent energy stability of the longest NVE run performed during this work is shown.
For details on the aging simulations see Sec.~\ref{sec:aging} below.

\clearpage

\section{Melting temperature}
It is commonly observed that ab-initio simulations based on DFT, and therefore also NNP trained on DFT data, exhibit melting temperatures $T_m$ of the liquid under study at variance with experiments \cite{malosso2022viscosity}. The usual procedure is to ``calibrate'' (i.e. shift) the simulation temperature by this difference, which then commonly results in good compatibility of simulation and experiments, for instance, concerning dynamical quantities.   
Thus, the melting temperature for toluene is determined here using the interface method, i.e., the crystalline phase is brought in contact with the liquid phase and simulations are performed in the NPT ensemble at different temperatures. If the chosen temperature is below $T_m$, the crystalline phase is supposed to grow, and for $T>T_m$ to melt. At or very close to $T_m$ the two phases can coexist for quite a long time. We performed extensive simulations on 244 molecules in the interface configuration as shown below. The lowest temperature at which we observed melting is $T_{\rm sim} = \SI{100}{K}$. $\SI{5}{K}$ below, phase coexistence was observed for the longest simulation time $t=\SI{240}{s}$ affordable. This corresponds to more than three times the relaxation time $\tau$ at this temperature. Neither a sign of the growth of the crystalline nor the liquid phase is observed after this time as can be seen in the snapshot and also from the radial distribution function below. Therefore, we believe that this temperature is close to $T_m$. As we do not observe crystallization at even lower temperatures, since the relaxation time exceeds the accessible simulation time, we can not give a lower error limit on $T_m$ from our simulations. As the experimental melting temperature is \SI{178}{K}, our result is

\begin{equation}
T_m^{\text{exp.}} - T_m^{\text{sim.}} = 80~K \pm {? K \atop 2 K}.
\end{equation}  

All temperatures from simulations reported in this work are shifted by this difference. \vspace{1cm}

\begin{center}
      t = 0\\
      \includegraphics[width=0.49\textwidth]{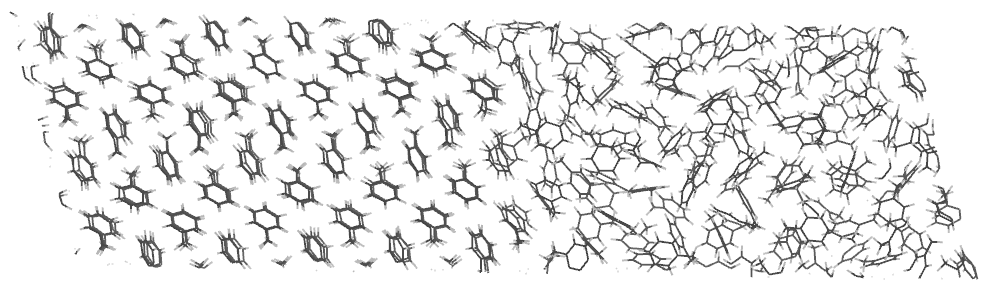}
  
    $T_{\rm sim}$ = 95 K, t = 240 ns \hspace{3cm } $T_{\rm sim}$ = 100 K, t = 240 ns\\
     \includegraphics[width=0.43\textwidth]{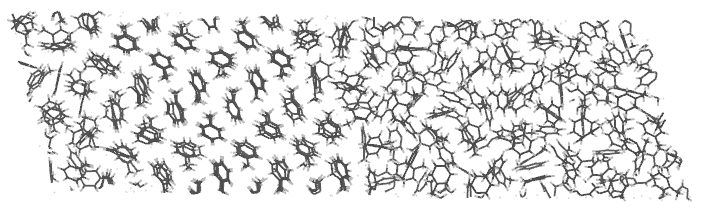}
    \hspace{0.5cm}\includegraphics[width=0.43\textwidth]{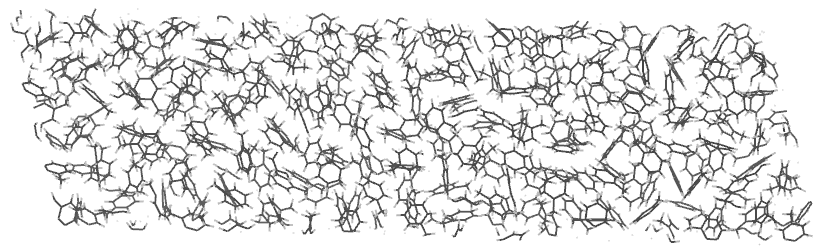}

    \includegraphics[width=0.75\textwidth]{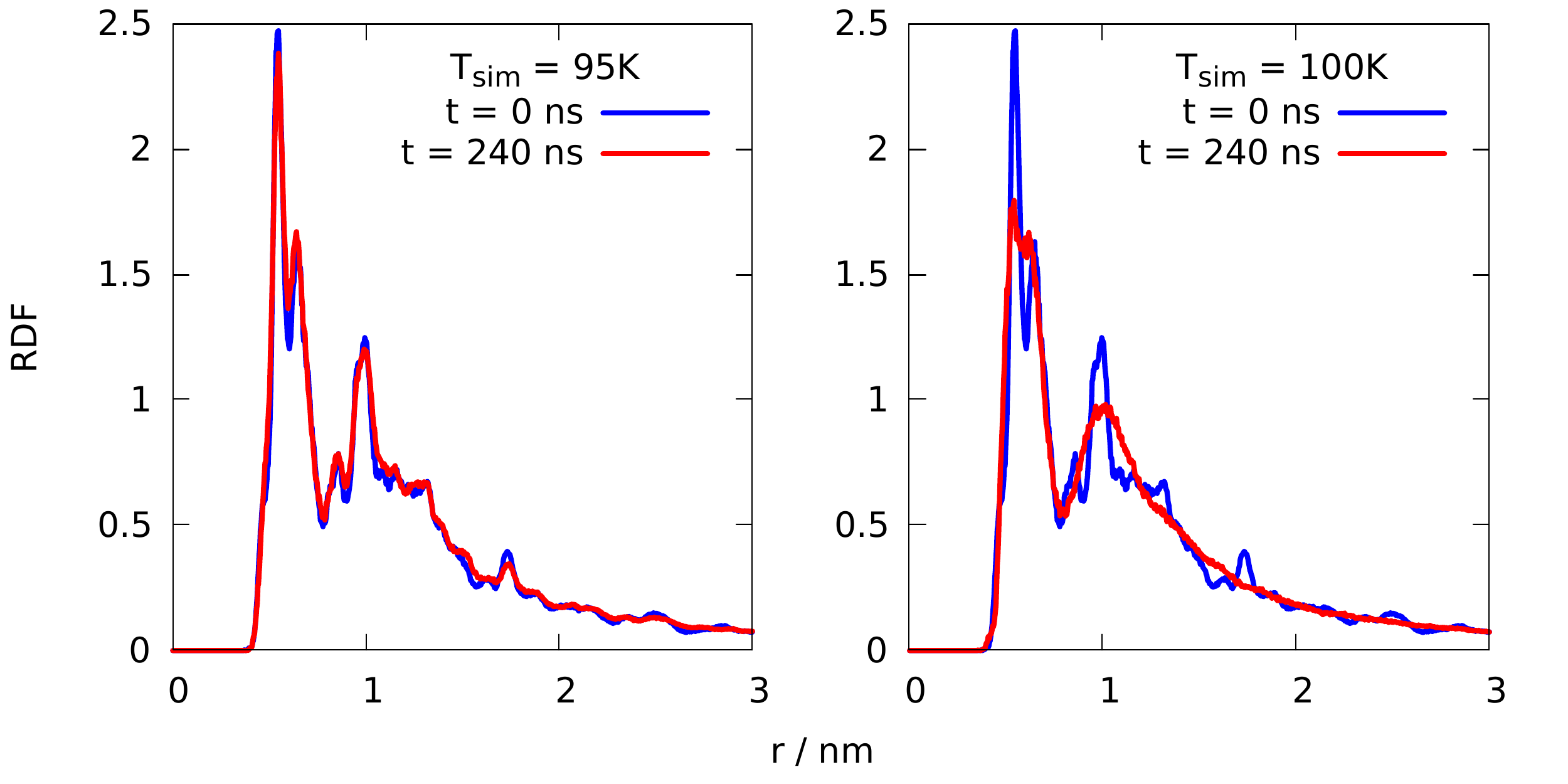}
\end{center}

%at 95K the corr time is 77.7ns -> interface sim time > 3$\tau$

\section{Overlap function}

\begin{figure}[h!]
    \centering
    \includegraphics[width=0.88\textwidth]{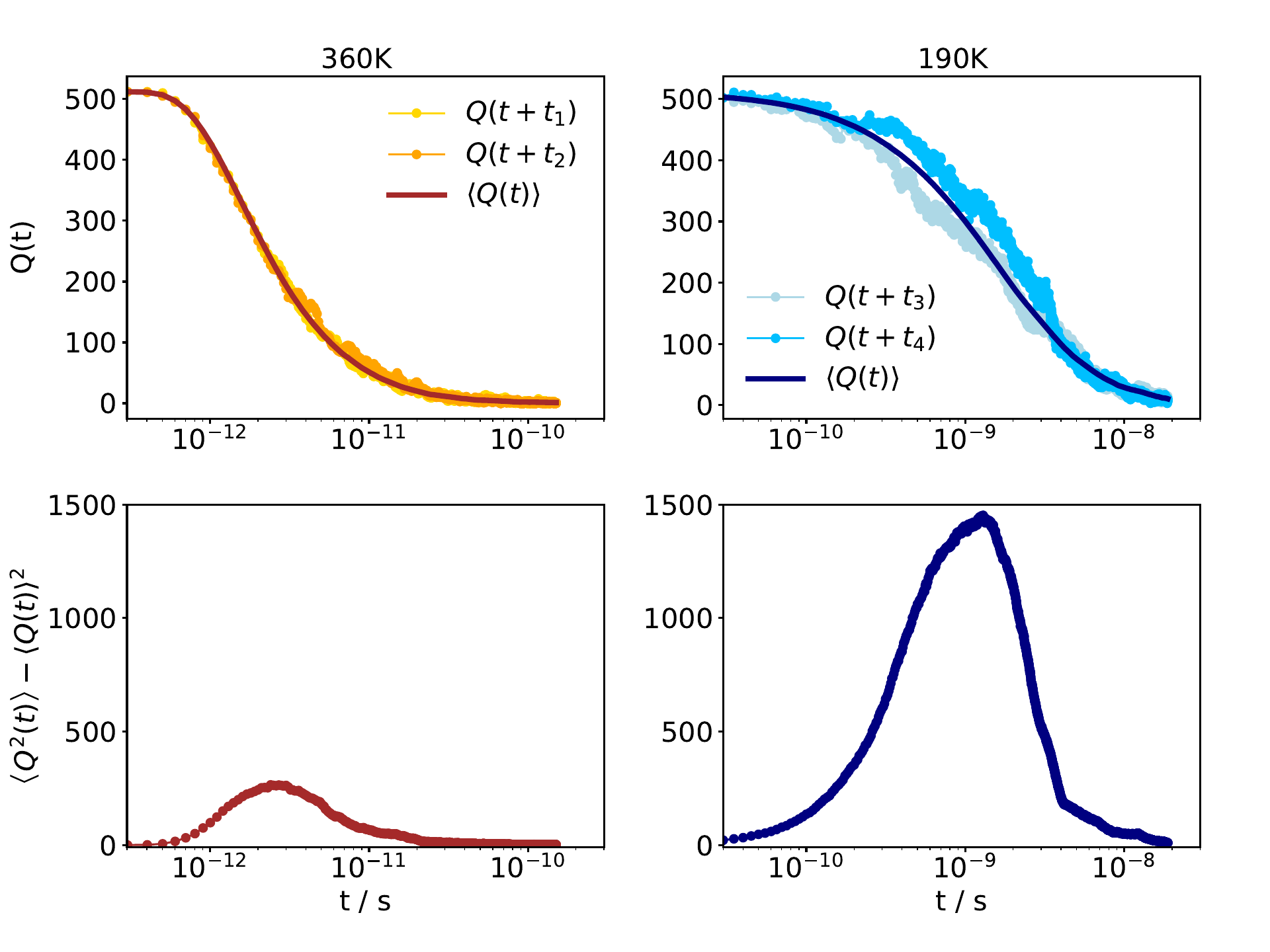}\
    \caption{Demonstration of how the difference in shape of $Q(t+t_i)$ at a high and a low temperature leads to the difference in the peak height of $\left< Q^2(t) \right> - \left< Q(t)\right>^2$. }
    \label{fig:overlap}
\end{figure}

In this section we illustrate how the different appearances of the overlap function $Q(t)$, defined in the main paper, lead to different peak heights of $\left< Q^2(t) \right> - \left< Q(t)\right>^2$ at different temperatures. The left column of Fig.~\ref{fig:overlap} shows data at \SI{360}{K} and the right column at \SI{190}{K}. In the top row, $Q(t+t_i)$ is shown for two randomly selected starting times $t_i$, as well as the average over all starting times $\left< Q(t) \right>$ at the respective temperature. It is obvious that at \SI{360}{K} the two $Q(t+t_i)$ are similar and also similar to the average, while great differences can be seen at intermediate times at \SI{190}{K}. Thus, it is immediately clear that the variance $\left< Q^2(t) \right> - \left< Q(t)\right>^2$ will be larger in this time range at the low temperature compared to the high temperature. This is shown in the lower row.

The reason behind the different appearances of the different $Q(t+t_i)$ at low temperatures is that a number of molecules move in a correlated fashion. This can be seen in the step-like shape of the $Q(t+t_i)$ at \SI{190}{K}. At some instances, the curve is approximately horizontal for a certain amount of time, while at another instance the curve suddenly drops, indicating the collective movement of a certain number of molecules at similar times. 
Since the sampling of these correlated motions depends on $t_i$, different $Q(t+t_i)$ will have a different shape leading to a higher variance the more molecules move in a correlated way.

\clearpage

\section{Viscosity} \label{sec:visc}
The viscosity of the system can be calculated using the Green-Kubo theory of linear response \cite{green1954markoff,kubo1957statistical}:
\begin{equation}
    \eta = \frac{V}{k_B T} \int\limits_0^{\infty} \left< \sigma_{\alpha\beta}(0)\sigma_{\alpha\beta}(t)\right> \text{d}t,
    \label{eq:GK}
\end{equation}

where $\sigma_{\alpha\beta}$ are the off-diagonal elements ($\sigma_{xy}$, $\sigma_{xz}$, $\sigma_{yz}$) of the stress tensor, over which is averaged in practice. Since the auto-correlation function $\left< \sigma_{\alpha\beta}(0)\sigma_{\alpha\beta}(t)\right>$ is very noisy, the variance of the integral grows linearly with the upper integration limit, making an error estimation difficult. To circumvent this problem, we used the open-source code SporTran \cite{ercole2022sportran}, which calculates the viscosity from the zero-frequency limit of the power spectrum $S(\omega)$ of the stress time series, i.e. , $S(\omega) = \int\limits_{-\infty}^{\infty} \left< \sigma_{\alpha\beta}(0)\sigma_{\alpha\beta}(t)\right> e^{i\omega t}\text{d}t$ and $\eta = \frac{V}{2 k_B T}S(0)$. 

\begin{figure}[h!]
    \centering
    \includegraphics[width=0.5\textwidth]{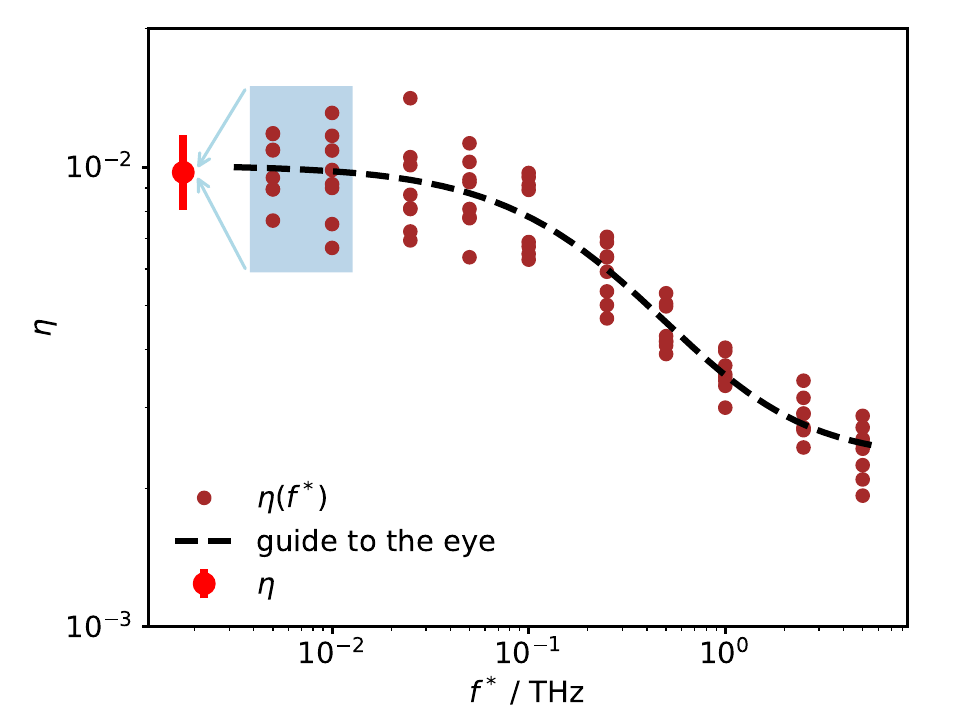}\
    \caption{Viscosity from 8 simulation runs calculated with different Nyquist frequency $f^*$. From the converged values (shaded area) the geometrical mean and its standard deviation is calculated and used as the final viscosity value at the respective temperature.  }
    \label{fig:sportran}
\end{figure}

For long trajectories it is necessary to adjust the cutoff (Nyquist) frequency $f^*$ used by the SporTran code to analyze the power spectrum. To find the optimal value of $f^*$ for each temperature, the viscosity is calculated using broadly varying values of $f^*$, and those values for which $\eta$ is converged with respect to $f^*$ are used. This procedure is exemplified for one temperature in Fig.~\ref{fig:sportran}, where the shaded area highlights the converged region from which the viscosity values of the 8 independent simulation runs are used to calculate the geometrical mean and its standard deviation. This value with the according error bar is taken as the final viscosity at the respective temperature. 

\clearpage

\section{Rotational correlation times} \label{sec:LS}
Rotational correlation times are usually determined by choosing a vector within the molecule, calculating its time auto-correlation function and taking the time for which it has decayed to $1/e$ as the rotational correlation time $\tau$.
Here, we chose a different route, which allows for direct comparison with experiments. Depolarized light scattering is an experimental method able to obtain rotational correlation functions for optical anisotropic molecules. It is sensitive to the off-diagonal elements of the polarizability tensor $\underline{\underline{\alpha}}$, which fluctuates over time mainly due to the rotational motions of the molecules \cite{pabst2022intensity}. Thus, we need to calculate 
\begin{equation}
    C(t) = \left<   \alpha_{ab}(0) \alpha_{ab}(t)  \right>
    \label{eq:corr}
\end{equation}
where $\alpha_{ab}$ are the off-diagonal elements ($\alpha_{xy}$, $\alpha_{xz}$, $\alpha_{yz}$) of the polarizabilty tensor of the whole simulation cell, over which is averaged in practice.
The polarizability tensor of the whole simulation cell is obtained with the phonon code of the \qe\ suite \cite{giannozzi2009quantum}. Due to the high computational cost of this procedure, it is impossible to get the polarizability tensor for large systems or for many time steps in this way. Therefore, we used only a total of 859 snapshots taken from the training set of the neural network potential described in Sec.~\ref{sec:NNP}, for which we calculated the polarizabilty tensor. With that, a neural network was trained to predict the polarizability tensor from the atomic coordinates using the tensor module of the DeepMD-kit \cite{zeng2023deepmd}.
In this way it is possible to obtain the polarizability tensor for all system sizes studied and for all time steps of the trajectories. $C(t)$ as obtained from from Eq.~\eqref{eq:corr} is shown for selected temperatures in Fig.~4 of the main paper. 

In order to compare the results from simulations directly to experimental data, we transformed $C(t)$ to the susceptibility representation in the frequency domain via Eq.~\ref{eq:FDT}, since this is the way experimental data on toluene is reported \cite{wiedersich2000comprehensive}:

    \begin{equation}
        \chi''(\omega) = \omega \int\limits_{-\infty}^{\infty} C(t) e^{i\omega t} \text{d}t
        \label{eq:FDT}
    \end{equation} 
    %Forster, Eq. 2.58 ($\hbar\to 0$, $S(\omega)=\int_{-\infty}^\infty C(t)e^{i\omega t}$, $C(t)$ real and even). 

\begin{figure}[h!]
    \centering
    \includegraphics[width=0.5\textwidth]{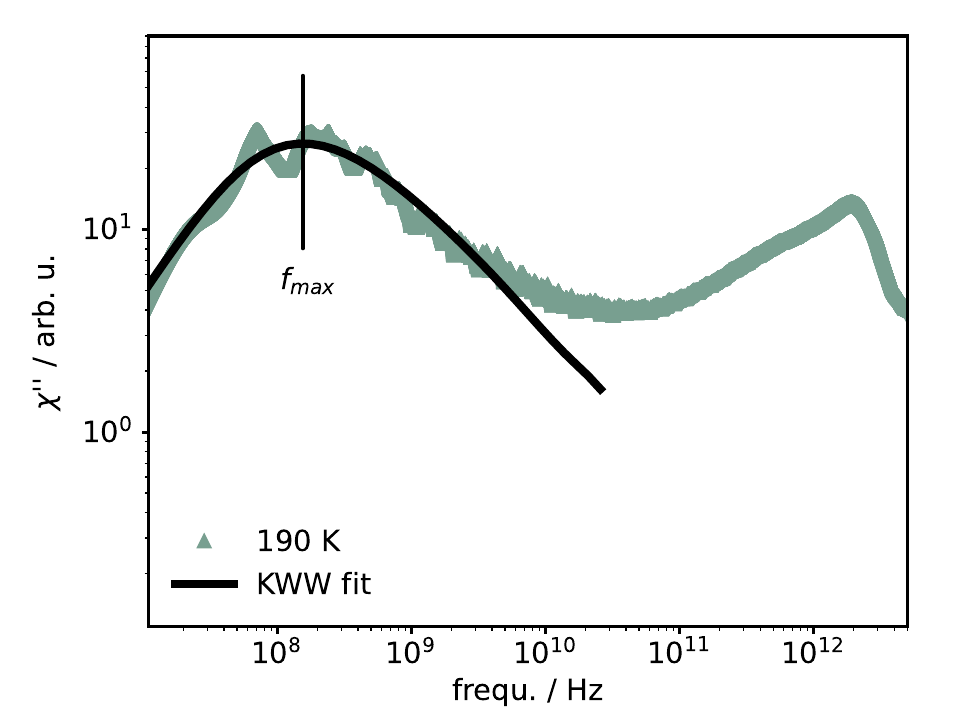}\
    \caption{Demonstration of the fitting procedure of the light scattering spectrum. The black solid line is the Fourier transform of a stretched exponential function. The frequency of its maximum is used to obtain the rotational correlation time $\tau$. }
    \label{fig:LS-fit}
\end{figure}

A comparison of a spectrum from experiment and simulation is given in Fig.~\ref{fig:visc-LS} below.
To obtain the rotational correlation time from these spectra, the same approach commonly used in experiments is employed: The structural relaxation peak of the $\chi''(\omega)$ spectrum is fitted with an empirical function, such as the stretched exponential $\exp(-(t/\tau)^{\beta})$, Fourier transformed to the frequency domain, and from the frequency of the peak maximum the rotational correlation time is obtained via $\tau = 1/(2\pi f_{max})$. This procedure is shown in Fig.~\ref{fig:LS-fit}. We note that we average the $C(t)$ from 8 independent simulation runs before using Eq.~\eqref{eq:FDT} and fitting.

\clearpage

\section{Comparison with Experimental Data}
\begin{figure}[h!]
    \centering
    \includegraphics[width=0.42\textwidth]{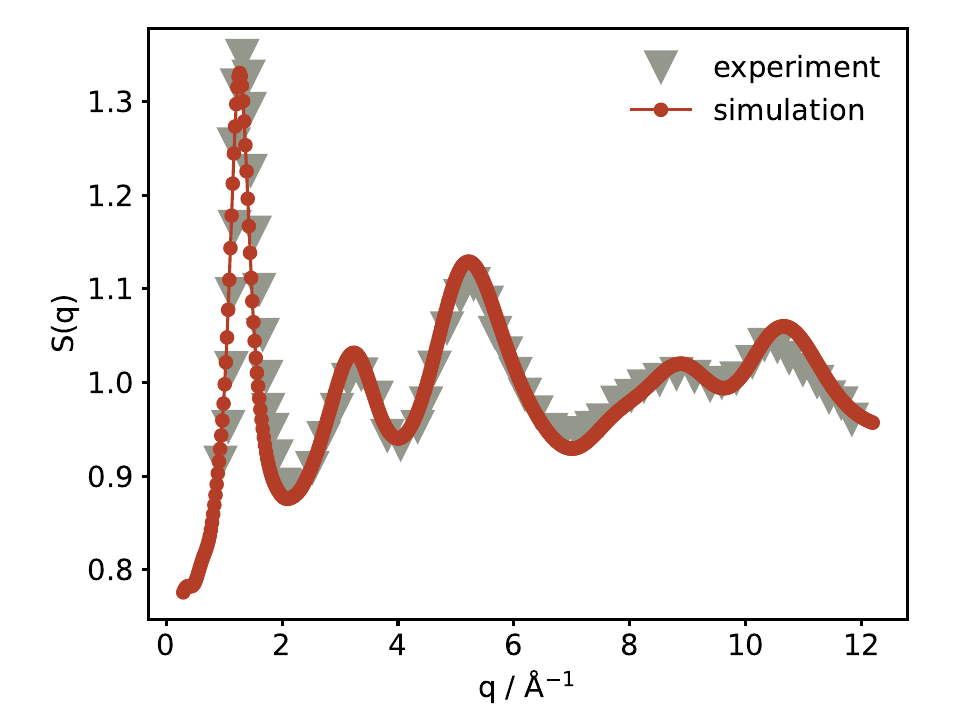}\hfill
    \includegraphics[width=0.42\textwidth]{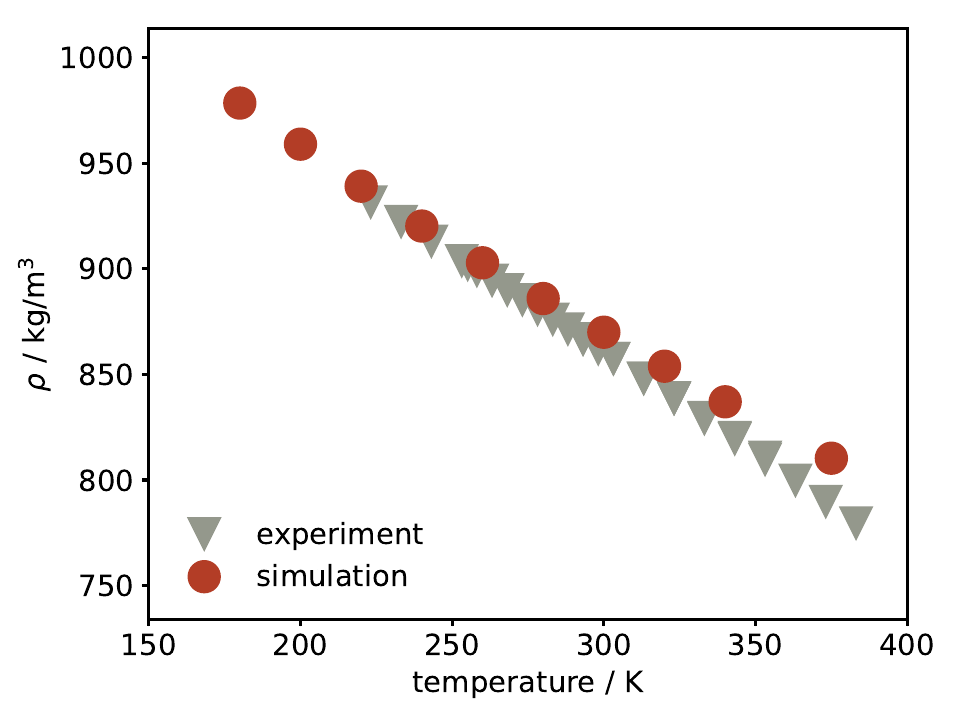}
    \caption{Structure factor $S(q)$ (left) and temperature dependent density $\rho$ (right) from experiment and simulations. }
    \label{fig:sfac-density}
\end{figure}

A comparison of experimental and simulation data is shown below for different static and dynamic quantities, highlighting the good accordance between simulation and experiment. On the left-hand side of Fig.~\ref{fig:sfac-density} the structure factor $S(q)$ is shown at ambient temperature. The experimental data is digitized from ref.~\citenum{headen2010structure}. The structure factor from the simulation is calculated using the TRAVIS code \cite{brehm2020travis}. Good agreement can be observed over the whole $q$-range.

On the right-hand side of Fig.~\ref{fig:sfac-density}, temperature dependent densities are shown. The experimental data is compiled from Refs.~\citenum{mclinden2008liquid,harris2000temperature,sanjun2007densities}. The data from simulations is obtained with 1024 molecules in the NPT ensemble after a sufficiently long equilibration time. At low temperatures simulation and experiment are in excellent agreement, while at higher temperatures approaching the boiling point slight deviations ($<3\%$) can be seen.

\begin{figure}[h!]
    \centering
    \includegraphics[width=0.42\textwidth]{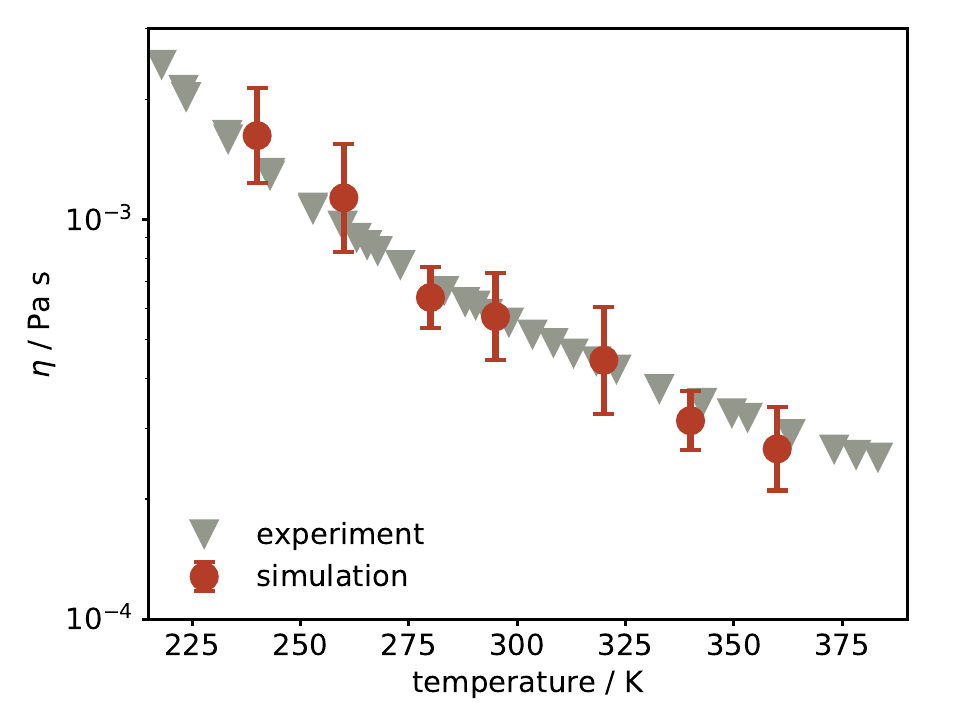}\hfill
    \includegraphics[width=0.42\textwidth]{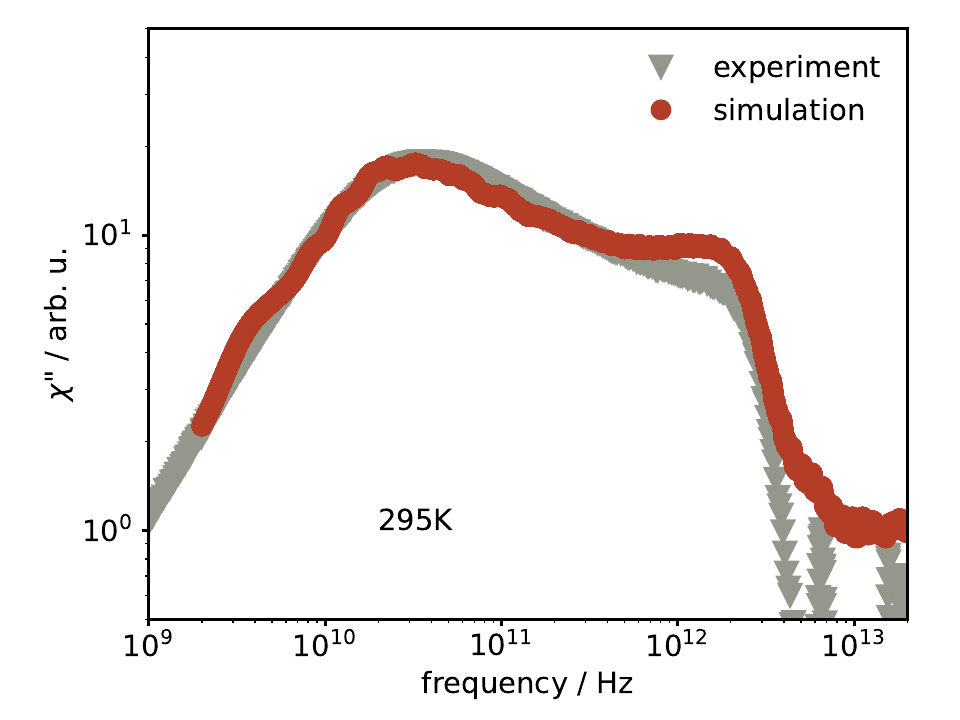}
    \caption{Temperature dependent viscosity (left) and depolarized light scattering spectrum (right) from experiment and simulation.}
    \label{fig:visc-LS}
\end{figure}

On the left-hand side of Fig.~\ref{fig:visc-LS} the temperature dependent viscosity $\eta$ is shown for not too low temperatures. Experimental data are from Refs.~\citenum{santos2006standard}, while the simulation data is the same as shown in Fig.~1 of the main paper, i.e., the one obtained on the 512 molecules system.   

On the right hand side of Fig.~\ref{fig:visc-LS}, the depolarized light scattering spectrum at \SI{295}{K} is shown. Experimental data is from ref.~\citenum{pabst2022understanding}. The way the simulated spectrum is obtained is detailed in Sec.~\ref{sec:LS}. Except for slight deviations in the vibrational band around \SI{2}{THz}, the agreement is very good.  

We note that the viscosity and rotational correlation times obtained from the simulations deviate at lower temperatures, starting approximately around \SI{220}{K}, from the experimental ones, in such way that they increase more quickly in the simulations. The reason for this is unclear at the moment and needs further study. However, this does not affect the results discussed in the main paper.

\clearpage

\section{Check for finite-size effects}

\begin{figure}[h!]
    \centering
    \includegraphics[width=0.48\textwidth]{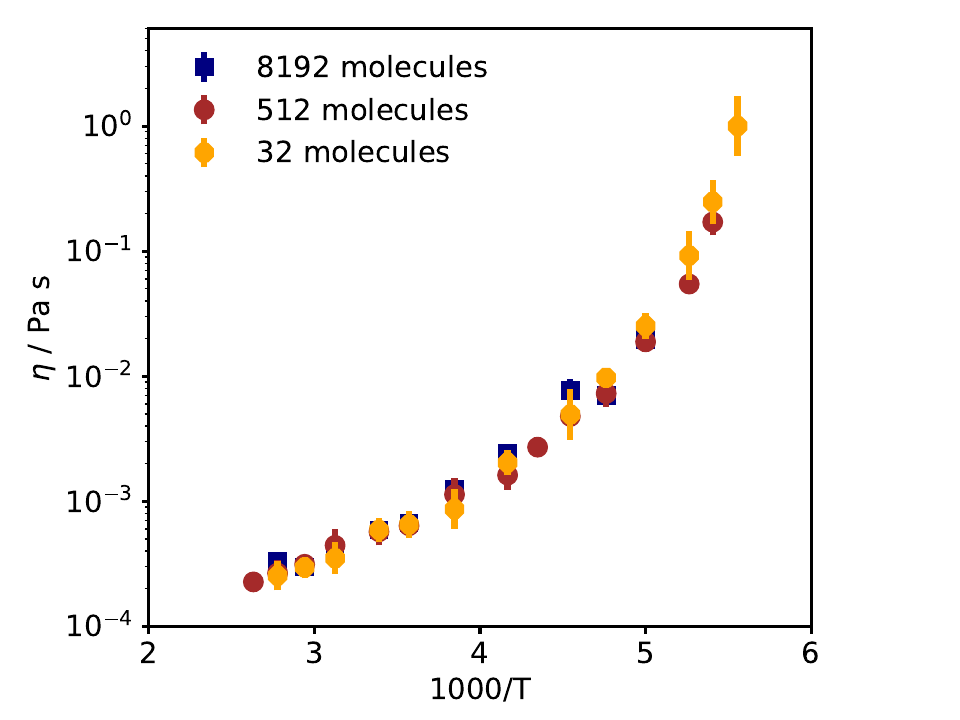}\hfill
    \includegraphics[width=0.48\textwidth]{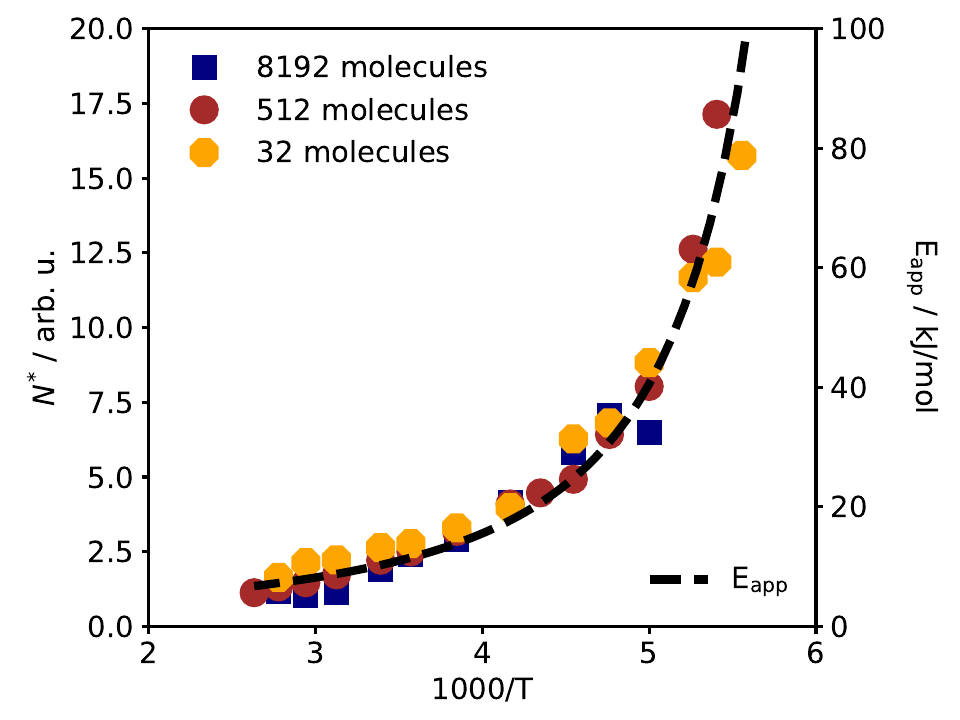}
    \caption{No finite size effects can be seen for the viscosity (left hand side) and the temperature dependence of the number of correlated molecules (right hand side) down to 32 molecules. }
    \label{fig:finite-size}
\end{figure}

Here, the results reported in Fig.~1 of the main paper are checked for possible influence of the finite size of the simulation box. To this end, we compare values obtained with simulation boxes containing 32 molecules (480 atoms), 512 molecules (7680 atoms), and 8192 molecules (122880 atoms).
On the left-hand side of Fig.~\ref{fig:finite-size} the viscosity is shown, which is calculated as detailed in Sec.~\ref{sec:visc}. Due to the high computational cost, only one simulation was run for the largest box (instead of 8 for the other box sizes). Thus, for the 8192 molecules box, the viscosity value is taken directly from one SporTran calculation with the error estimate from the same code. The longest simulations could only be run for the smaller boxes, but in the range where the viscosity value from more than one box size is available, good agreement within the error bars can be seen, meaning that no finite-size effects are influencing the viscosity down to 32 molecules. 

On the right-hand side of Fig.~\ref{fig:finite-size}, the values of the number of correlated molecules $N^*$, are shown for different box sizes. We note that the absolute magnitude of $N^*$, depends slightly on the box size for unknown reasons. However, the temperature dependence of the values, which is the important part, is identical for the different box sizes as seen from the data, which was slightly shifted on the y-axis by factors such that the data for different box sizes overlap.

\clearpage

\section{Structure-Dynamics interplay}
\begin{figure}[h!]
    \centering
    \includegraphics[width=0.76\textwidth]{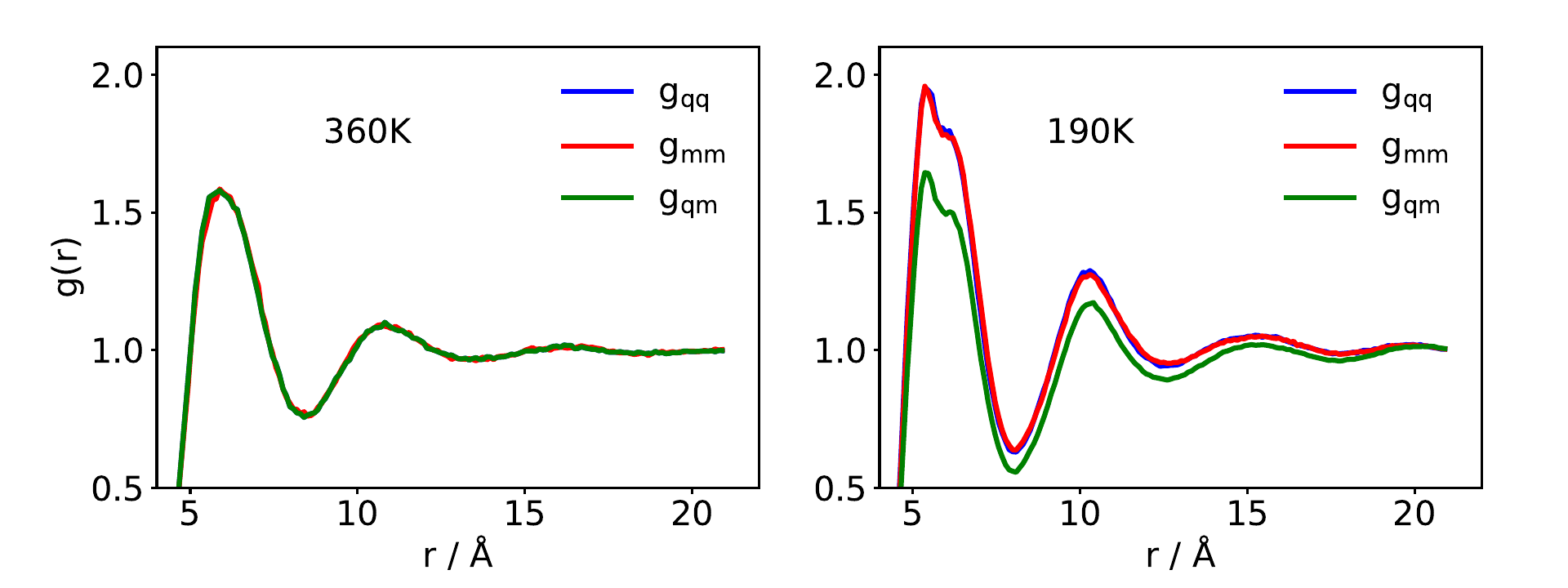}
    \caption{Partial radial distribution functions at a two different temperatures, where $q$ stands for quiescent and $m$ for mobile.}
    \label{fig:gr}
\end{figure}

The spatial dynamic heterogeneity shown in Fig.~2 of the main paper can also be visualized using the partial radial distribution functions $g_{xx}(r)$, where $xx$ is either $mm$, i.e., the partial $g(r)$ of only the mobile molecules, or $qq$ for the quiescent molecules. The third case is the distinct part $qm$, which monitors the probability of finding a mobile molecule at $r$ from a quiescent molecule, or vice versa. These partial radial distribution functions are shown for a high and low temperature in Fig.~\ref{fig:gr}. While at \SI{360}{K} all three $g_{xx}(r)$ are basically identical, a pronounced difference between $g_{qm}(r)$ and $g_{mm}(r)$, $g_{qq}(r)$ can be seen at \SI{190}{K}, with the former being markedly reduced compared to the latter two at small $r$. The information contained in this observation is the same as in the lower part of Fig.~2 of the main paper: It is more probable at low temperatures to find molecules with the same mobility close to each other than with disparate mobility, while the distribution is random at high temperatures. \newline

More subtle effects of the structure-dynamics interplay can be visualized by utilizing a method proposed recently in ref.~\citenum{zhang2020revealing} to obtain the 3D structure of a liquid. In short, this method works as follows: A local coordinate system is build by using the position of a molecules (here: its centre of mass) as the origin, the direction to its nearest neighbor as the z-axis and the plane containing the second nearest neighbor as the x-z-plane. In this way a spherical coordinate system $\theta,\phi,r$ can be defined, which is used to measure the probability of finding a particle at a given point in space.  

\begin{figure}[h!]
    \centering
    \includegraphics[width=0.76\textwidth]{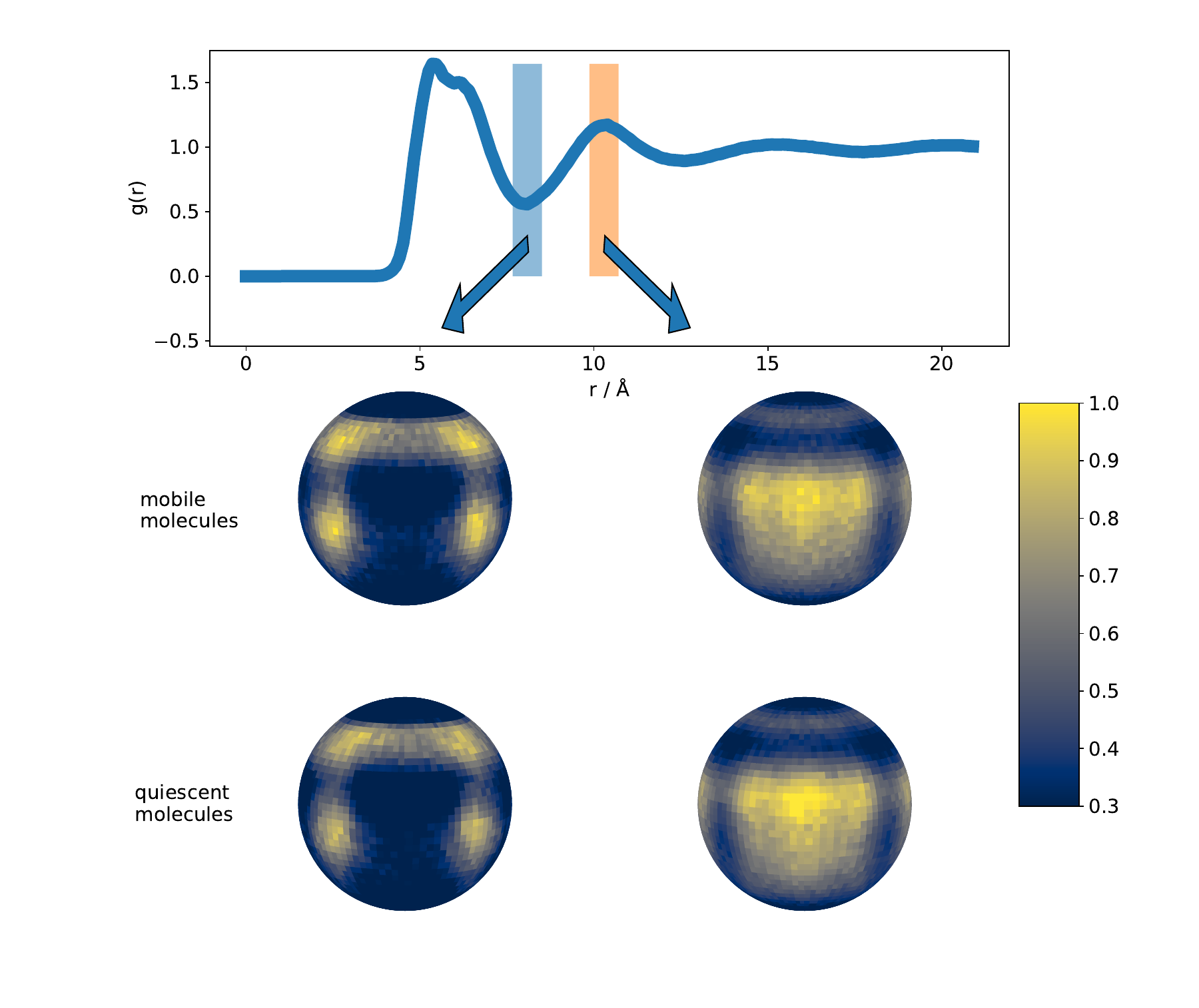}
    \caption{3D resolved probability of finding mobile (quiescent) molecules next to a mobile (quiescent) molecule at a distance corresponding to the first minimum of $g(r)$ (left-hand side) or to the second maximum (right-hand side).  }
    \label{fig:3D}
\end{figure}

Here, we combine this information about the 3D structure with the dynamical information of quiescent and mobile molecules (as defined in the main paper). The result is shown for \SI{190}{K} in Fig.~\ref{fig:3D}, where in the upper panel the total radial distribution function $g(r)$ is shown. The shaded areas highlight two regions for which the 3D structure in combination with the mobility information is shown in the spherical plots below. On the left-hand side $r$-values corresponding to the first dip in $g(r)$ and on the right-hand side to the second peak in $g(r)$ are shown. For each side, the colors represent the number of molecules found in this $r$-range at angles $\theta,\phi$, normalized between 0 and 1, i.e., the brighter a pixel in these spherical plots, the more probable to find a molecule at this $r,\theta,\phi$ combination. The two rows distinguish between the different mobilites of the molecules. In the upper row mobile molecules surrounding a mobile molecule located at the coordinate origin are counted. The same is done for quiescent molecules in the lower row. 
It can be seen from the brightness of the yellow spots in the left column that it is more likely to find mobile molecules  at distances from the reference molecule corresponding to the first dip in the radial distribution function than to find quiescent molecules. The opposite is true for distances corresponding to the second peak in $g(r)$, as shown in the right column.
In fact, is is quite intuitive that in regions which are populated by less molecules than average, mobility might be higher than in regions with high population. 

As already mentioned in the main paper, these findings highlight the fact that the spatial dynamic heterogeneities can not be viewed as rigid regions in space, within which necessarily all molecules move cooperatively. Instead, the dynamic correlations observed with decreasing temperature exhibit subtle variations coupled to the structure of the liquid, reaching from the nearest neighbors to distances at of least two inter-particle distances.

\clearpage

\section{Aging}\label{sec:aging}
\begin{figure}[h!]
    \centering
    \includegraphics[width=0.46\textwidth]{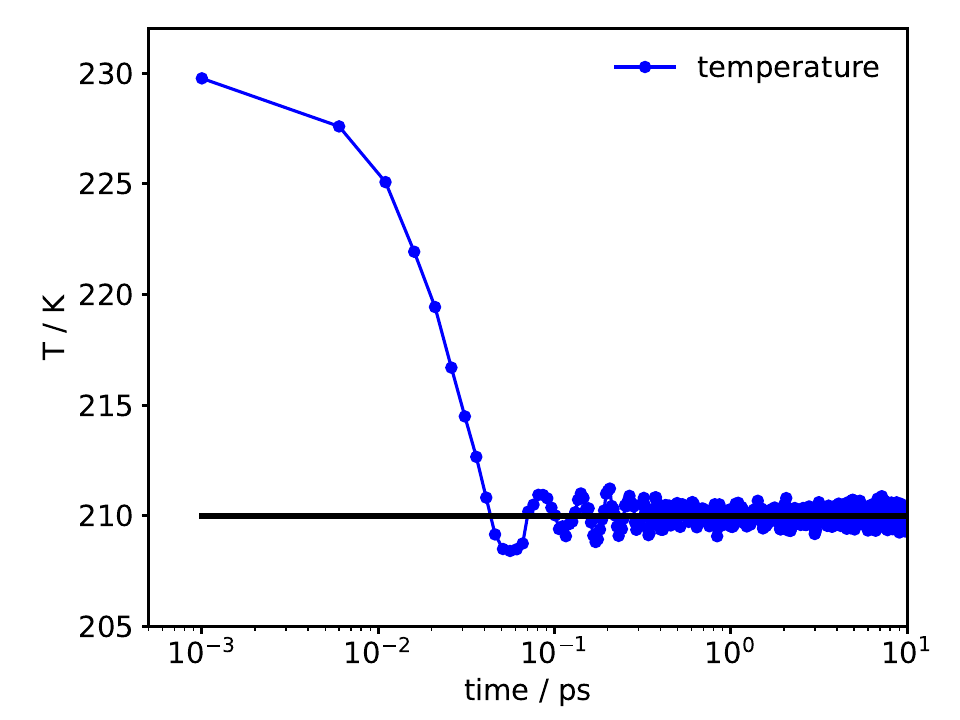}\hfill
    \includegraphics[width=0.46\textwidth]{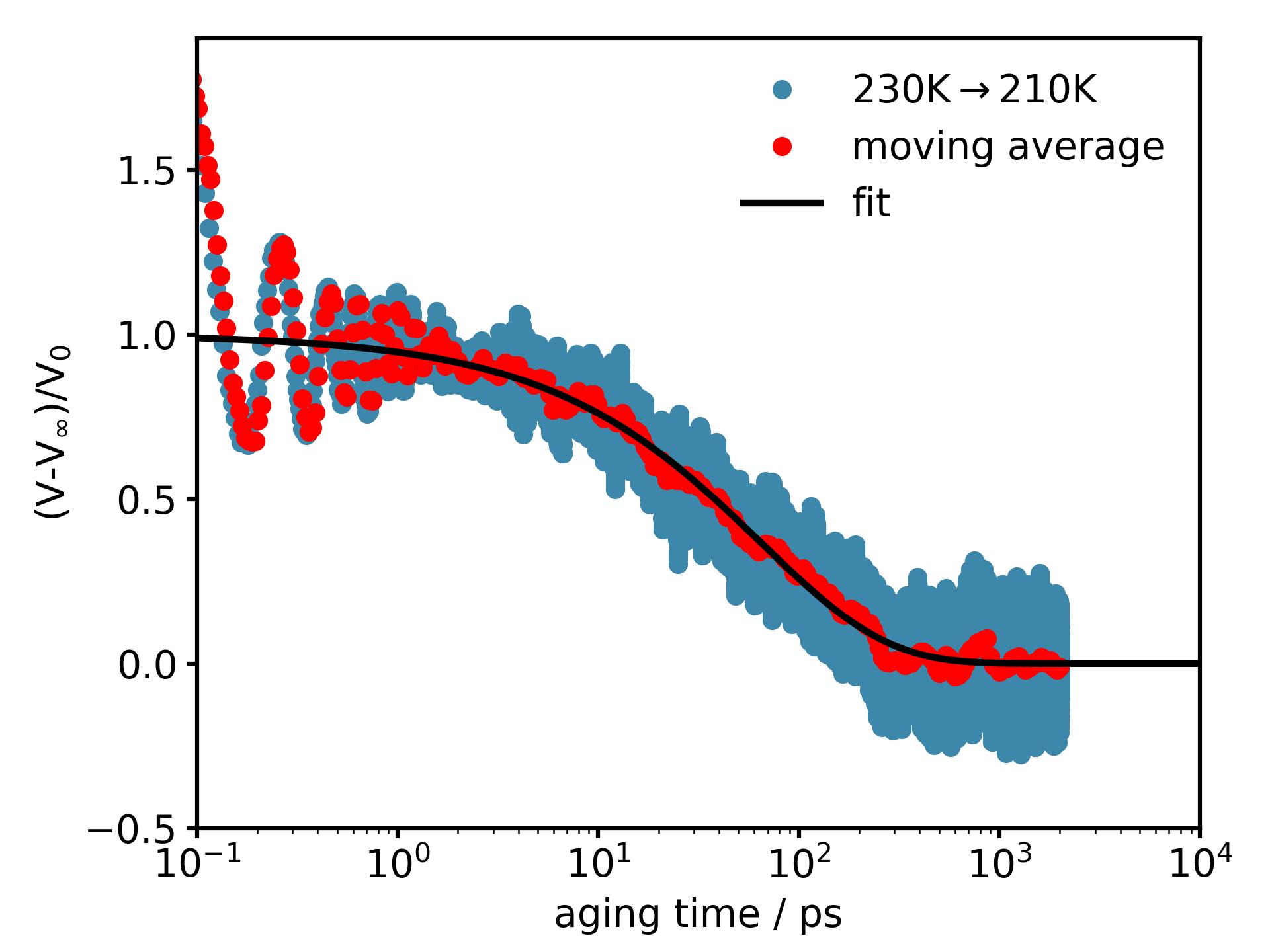}
    \caption{Left-hand side: Temperature evolution of after the \SI{20}{K} down-jump. Right-hand side: Comparison of the volume data after the \SI{20}{K} down-jump with its logarithmic moving average. }
    \label{fig:aging}
\end{figure}

Aging simulations were performed in the NPT ensemble on 512 molecules, starting from well equilibrated runs at \SI{230}{K} and \SI{190}{K} jumping to \SI{210}{K}. In total 20 simulations starting from independent points of the equilibration run were performed for both up- and down-jump, over which was averaged. The time step for all aging simulations was set to \SI{0.5}{fs}.
To arrive at the final temperature as quickly as possible, the damping parameter of the thermostat was initially set to \SI{10}{fs} and after a simulation time of \SI{1}{ps} increased to \SI{50}{fs} in order to minimize the influence of the thermostat on the dynamics. This procedure is inspired by ref.~\citenum{jaeger2022temperature}. On the left-hand side of Fig.~\ref{fig:aging}, the temperature evolution of the system after a temperature down jump is shown. It can be seen that the temperature is stable at the final temperature on a sub-picosecond time scale, ensuring that the aging event can be followed starting from the shortest times.

On the right-hand side, the comparison of the volume evolution (average over 20 independent runs, blue points) and its moving average (on a logarithmic time scale, red points) reported in the main part of the paper is shown. Additionally, a fit to the data using a stretched exponential decay is shown as a black line. From this fit it got clear that the time scale of aging is in very good agreement with the rotational correlation time obtained in equilibrium, as it is expected.  

The value of $N^*$ is evaluated during aging like it is detailed in the main paper with one important difference: Since the liquid is not in equilibrium during aging, the variance $\Delta Q^2(t)$ can not be evaluated using different starting times $t_0$ to calculate $Q(t)$. Instead, $Q(t)$ is calculated for one fixed aging time for each of the 20 independent aging runs and the variance between these runs is used to calculate $N^*$. This procedure was used before in the literature \cite{parsaeian2008growth}. 
Due to the small number of aging runs, the statistics is way worse than in equilibrium, leading to noisy data. For unknown reasons, the data for the temperature up-jump is even more noisy than the down-jump (and therefore not shown in the main paper), and due to the high computational costs of these calculations we leave it for future studies to improve the statistics by performing more runs.

\clearpage

\section{Time-Temperature-Superposition}
\begin{figure}[h!]
    \centering
    \includegraphics[width=0.46\textwidth]{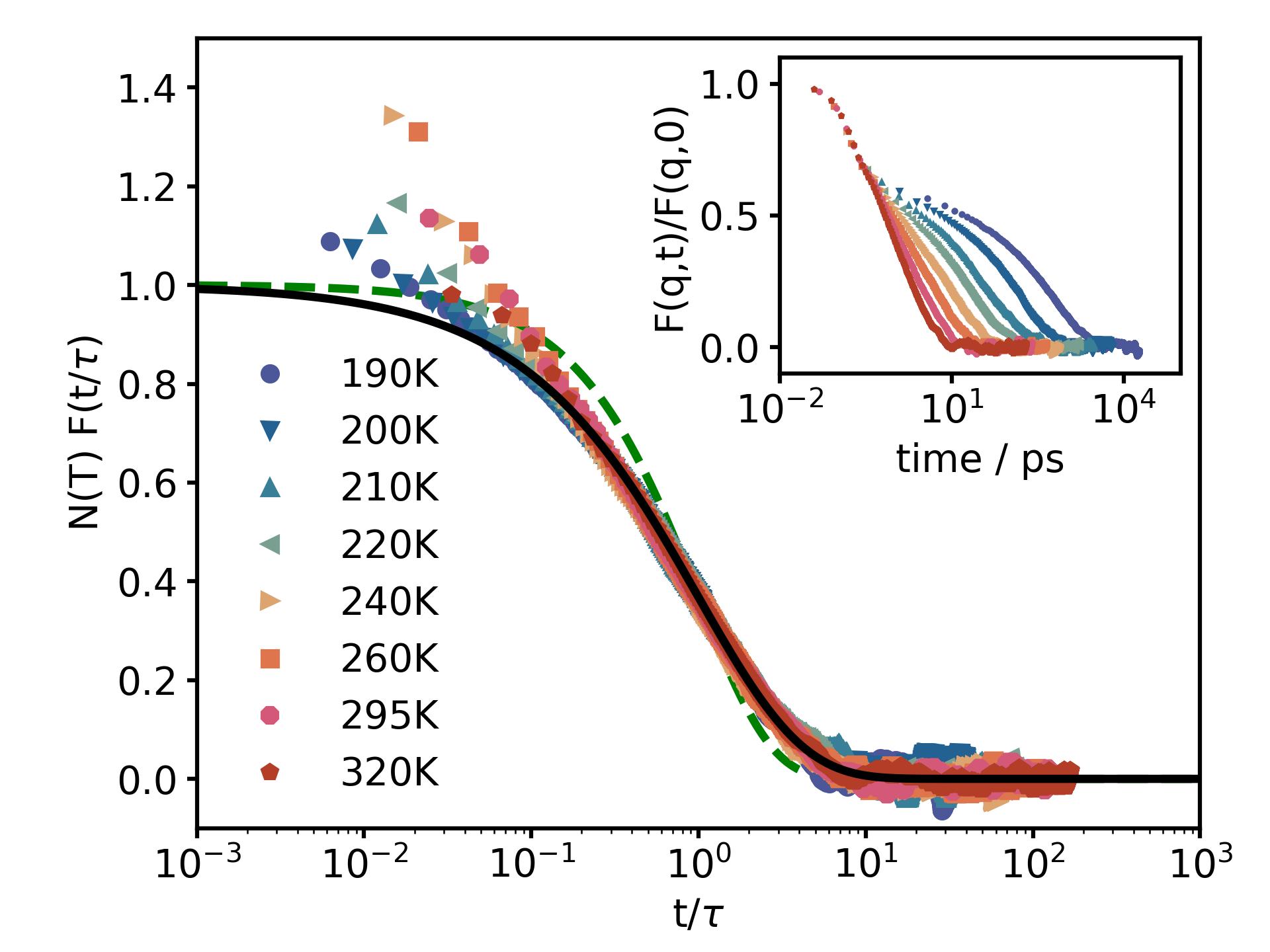}\hfill
    \caption{Test for time-temperature superposition of the intermediate scattering function. Black solid line is a stretched exponential decay, while the dashed line is an exponential decay. }
    \label{fig:fqt}
\end{figure}

In the main part of the paper, time-temperature-superposition (TTS), i.e., the invariance of the shape of correlation function with repect to temperature, is reported for the rotational correlation function. The same is shown here for translational dynamics as probed by the intermediate scattering function
\begin{equation}
  F(\bm q,t) = \frac{1}{N} \left< \sum\limits_{i=1}^N \sum\limits_{j=1}^N \exp\left(-i\bm q (\bm r_i(t+t_0) - \bm r_j(t_0))\right) \right>  ,
\end{equation}
where $\bm q$ is chosen close to the first peak in the structure factor $S(\vec q)$. 
In the inset of Fig.~\ref{fig:fqt}, $F(t)$ is shown normalized between 0 and 1 for selected temperatures as calculated with the LiquidLib code~\cite{walter2018liquidlib}.
In the main panel, the same data is superimposed by shifting it horizontally by the correlation time $\tau(T)$ and vertically by a factor N(T), such that best overlap of the curves is obtained.  

It can be seen from the superposition that the shape of the intermediate scattering function is temperature independent to good approximation. Of course, at high temperatures, differences show up as the structural relaxation is cut off at short times by vibrational motions. The black solid line is a stretched exponential with $\beta=0.7$ and the dashed line an exponential decay, i.e., $\beta=1$. The same conclusions as for the rotational dynamics in the main paper can here be drawn for the translational motions: An exponential decay is not observed even at temperatures close to the boiling point. The shape of the structural relaxation is independent of temperature, despite the fact that dynamical heterogeneity increases with decreasing temperature. 

\clearpage

\section{Stokes-Einstein relation}
\begin{figure}[h!]
    \centering
    \includegraphics[width=0.46\textwidth]{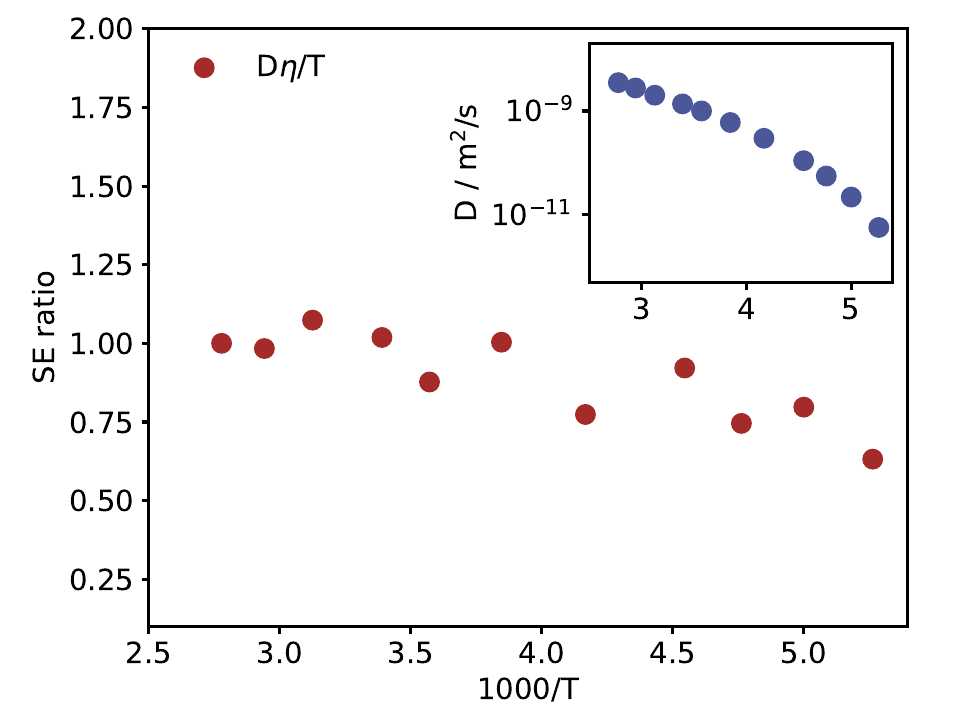}\hfill
    \caption{The Stokes-Einstein ratio, not showing increasing values at low temperatures, as commonly assumed to be caused by dynamical heterogeneity.}
    \label{fig:SE}
\end{figure}

A violation of the Stokes-Einstein (SE) relation $D(T)\eta(T)/T =$ const. is often reported in the literature for liquids at low temperature. The observation is that upon lowering the temperature, the viscosity $\eta(T)$ increases more strongly than the diffusion constant $D(T)$ decreases, leading to an increase in the SE-ratio. It is often assumed that this effect is due to dynamic heterogeneity. The idea behind this is that the diffusion constant is dominated by the fast particles, while the viscosity is not. In the inset of Fig.~\ref{fig:SE} the diffusion constant D as obtained from the mean square displacement with the TRAVIS code \cite{brehm2020travis} is shown over inverse temperature. The SE-ratio is shown in the main panel of this Fig.. It can be seen that the SE-ratio is nearly constant upon decreasing temperature. If at all, it slightly decreases at lower temperatures. Since the dynamical heterogeneity markedly increases in this temperature range, as shown in the main part of the paper, we can conclude that an increase in dynamical heterogeneity does not inevitable lead to a violation of the SE relation.

\bibliography{bib}% Produces the bibliography via BibTeX.